\documentclass[aps,pre,twocolumn,a4paper,superscriptaddress,floatfix,10pt,longbibliography]{revtex4-2}

\usepackage[T1]{fontenc}
\usepackage{graphicx}
\usepackage{dcolumn}
\usepackage{bm}
\usepackage{color}
\usepackage{amsmath}
\usepackage{amssymb}
\usepackage{hyperref}
\usepackage{subfigure}
\usepackage{float}
\usepackage{bigints}
\usepackage{changes}
\usepackage{braket}
\usepackage{xcolor}

\newcommand{\gs}{{\mathrm{gs}}}
\newcommand{\mm}{{\mathrm{min}}}

\newcommand{\U}{{\hat{U}}}
\newcommand{\X}{{\hat{X}}}

\newcommand{\Z}{{\hat{Z}}}
\newcommand{\tr}{{\mathrm{tr}}}
\newcommand{\ts}{{\mathrm{ts}}}

\begin{document}

\title{Deep learning non-local and scalable energy functionals for quantum Ising models}

\author{E. Costa}
\affiliation{School of Science and Technology, Physics Division, Universit\`a di Camerino, 62032 Camerino, Italy}
\affiliation{INFN-Sezione di Perugia, 06123 Perugia, Italy}

\author{R. Fazio}
\affiliation{Abdus Salam ICTP, Strada Costiera 11, I-34151 Trieste, Italy}
\affiliation{Dipartimento di Fisica, Universit\`a di Napoli ``Federico II'', Monte S. Angelo, I-80126 Napoli, Italy}

\author{S. Pilati}
\affiliation{School of Science and Technology, Physics Division, Universit\`a di Camerino, 62032 Camerino, Italy}
\affiliation{INFN-Sezione di Perugia, 06123 Perugia, Italy}

\begin{abstract}
Density functional theory (DFT) is routinely employed in material science and in quantum chemistry to simulate weakly  correlated electronic systems. Recently, deep learning (DL) techniques have been adopted to develop promising functionals for the strongly correlated regime. DFT can be applied to quantum spin models too, but functionals based on DL have not been developed yet.
Here, we investigate DL-based DFTs for random quantum Ising chains, both with nearest-neighbor and up to next-nearest neighbor couplings. Our neural functionals are trained and tested on data produced via the Jordan-Wigner transformation, exact diagonalization, and tensor-network methods. An economical gradient-descent optimization is used to find the ground-state properties of previously unseen Hamiltonian instances. Notably, our non-local functionals drastically improve upon the common local density approximations, and they are designed to be scalable, allowing us to simulate chain sizes much larger than those used for training. The prediction accuracy is analyzed, paying attention to the spatial correlations of the learned functionals and to the role of quantum criticality. 
Our findings indicate a suitable strategy to extend the reach of other computational methods with a controllable accuracy.
\end{abstract}

\maketitle

\section{Introduction}
Density functional theory (DFT) is the workhorse of computational material science and of theoretical quantum chemistry~\cite{perspective_dft}. It is routinely and successfully employed to perform electronic structure simulations. The predictions provided by the known approximations for the (unknown) energy-density functional are often accurate. However, dramatic failures often occur when the electrons are strongly correlated~\cite{cohen2012challenges}. 
In the last few years, deep learning (DL) techniques have started pervading physics research~\cite{Dunjko_2018,carleo2019machine,doi:10.1080/23746149.2020.1797528}, including the field of electronic-structure theory~\cite{Kulik_2022}. Indeed, neural networks have already been used to develop functionals from data~\cite{bypassing_kohm_sham,Li_machine_learning_dft,li_machine_learning_two,doi:10.1126/science.abj6511,machine_learning_hubbard}, providing a promising strategy to tackle strongly-correlated systems. Incidentally, deep learning also facilitates implementing orbital-free DFT~\cite{orbital_free_ryczko_voxel_cnn,Meyer2020,snyder_rupp_machine_learning,costa_scriva}, allowing simulating ground states with the computationally convenient gradient-descent optimization.

Notably, DFT can be applied also to Hamiltonians defined using countable local bases~\cite{lattice_dft,dft_on_graphs}, including quantum spin models~\cite{dft_heisenberg_model,dft_ising_chain}. For the latter systems, it was proven that, in principle, the ground-state energy can be determined from arrays of the magnetizations on all sites. 
In recent years, quantum spin Hamiltonians have received increased attention, since they quantitatively describe important quantum-simulation platforms such as Rydberg atoms in optical tweezers~\cite{optical_tweezers,doi:10.1126/science.aax9743}, trapped ions~\cite{Smith2016,Morong2021}, quantum annealers based on superconducting qubits~\cite{boixo2014evidence}, photonic circuits~\cite{pitsios2017photonic}, as well as various setups of cold-atoms in optical lattices (see, e.g., Refs~\cite{PhysRevLett.115.215301,jepsen2020spin}).
These systems are reaching regimes --  in terms of lattice size, dimensionality, interaction range, and frustration -- where common computational methods, e.g., tensor-network methods, become computationally prohibitive. 
DFT represents a suitable alternative. Unfortunately, magnetization functionals for quantum spin models have not been investigated as thoroughly as density functionals for electronic systems. In particular, deep learning techniques have not yet been adopted yet.

In this Article, we investigate DFTs for quantum Ising models using scalable deep neural networks.
Our testbeds are Ising chains with random transverse field, either with nearest-neighbor interactions only, or including up to next nearest-neighbor couplings.
Our main goal is to show that accurate magnetization-energy functionals can be developed using deep neural networks trained on small, computationally feasible, chain sizes. 
As we demonstrate, the ground-state energies and magnetization profile of previously unseen random Hamiltonian instances can be determined via a stable and computationally convenient gradient-descent optimization. 
An analysis of the gradient-descent error is provided, considering in particular the ferromagnetic quantum critical point, showing the role of the disorder strength and the systematic control of the prediction accuracy.
Notably, the architecture of our networks is designed to be scalable. This allows simulating chain sizes much larger than those used during the training of the functional. 
Special attention is paid to the role of the chain size used for training. It is found that, in the worst-case scenario corresponding to the quantum critical point, the gradient-descent error in the asymptotic large-size regime decreases approximately with the third power of the training chain length. An analysis of the input-output statistical spatial correlations of local observables is also provided. This analysis suggests a heuristic model that describes the observed scaling of the prediction accuracy with the training size.

The rest of the Article is organized as follows:
Section~\ref{sectionmethods} introduces the required notions on DFT for discrete systems.
The testbed Hamiltonians we consider are described in Section~\ref{section: hamiltonians}.
Our DL approach to DFT for random quantum Ising models is described in Section~\ref{secdldft}.
The accuracy of the DL-DFT approach in the testbed mentioned above is analyzed in Section~\ref{Results}.
Section~\ref{secconclusions} summarizes our main findings and mentions some future perspectives.
Appendix~\ref{section: systematic improvement} reports additional results to demonstrate the systematic control of the DL-DFT accuracy.
In Appendix~\ref{section: binder cumulant}, we determine the ferromagnetic quantum critical point of the quantum Ising chain including also next-nearest neighbor interactions.

\begin{figure}[H]
\centering
\includegraphics[width=1.\columnwidth]{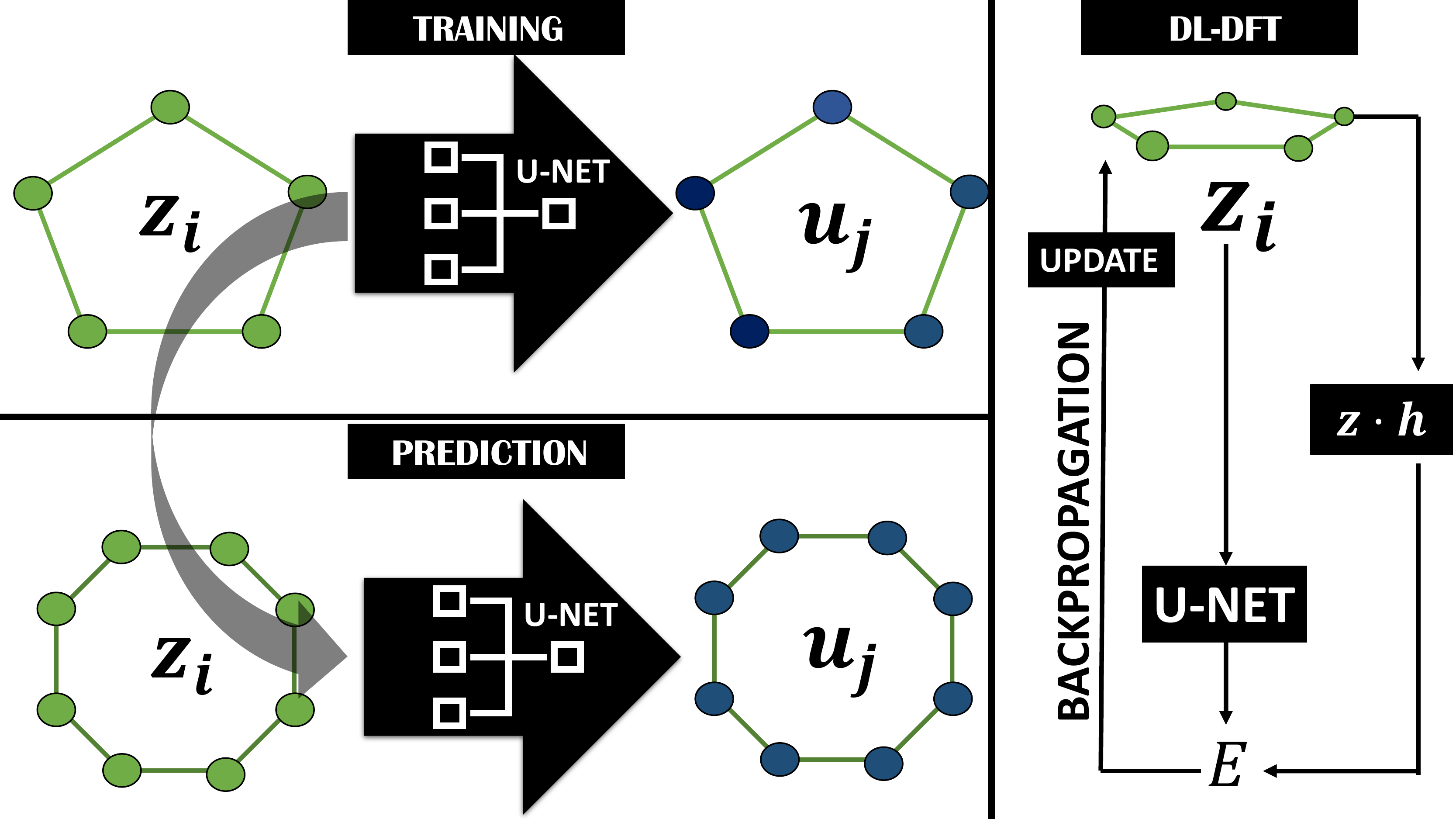}
\caption{Schematic representation of our DL-DFT approach.
The training is performed on small lattices, and it allows a convolutional neural network (U-NET) learning the mapping between the magnetizations $z_i$ and the energy terms $u_j$. The trained U-NET can be used to predict the energy terms for larger lattice sizes. The ground-state energy and magnetization profile are found by gradient-descent minimization of the DL-DFT energy per spin $E$. The energy gradients are computed exploiting the backpropagation algorithm implemented in PyTorch~\cite{NEURIPS2019_9015}.
}
\label{vs_training_size}
\end{figure}

\section{Formalism of density functional theory}
\label{sectionmethods}
DFT is an efficient computationally method commonly used to simulate the ground state of continuous-space Hamiltonians describing electronic systems~\cite{perspective_dft}.
The first Hohenberg-Kohn theorem states that the ground-state energy of (non-degenerate) Hamiltonians with fixed interactions and variable external potentials $v(x)$ is a  functional of the particle density $n(x)$~\cite{h-k_paper}. Together with the variational property ensured by the second Hohenberg-Kohn theorem, this allows searching for the ground-state energy by minimizing a universal energy functional of the density $E[n(x)]$~\cite{Levy_theory_dft}.
It is worth mentioning that in the formulations of DFT by Levy~\cite{Levy_theory_dft} and Lieb~\cite{Lieb2002} some constraints on the allowed densities $n(x)$ have been overcome, and degenerate ground-state can be accounted for.
DFT can be applied also to discrete-variable Hamiltonians, including both lattice~\cite{dft_on_graphs} and spin models~\cite{dft_ising_chain,dft_heisenberg_model}.
Rather generally, the Hamiltonian must include a fixed term $\U$, and a discrete set of local operators $\hat{\mathbf{O}}=(\hat{O}_1,\hat{O}_2,\dots)$, coupled to just as many variable fields $\mathbf{h}=(h_1,h_2,\dots)$. Basically, the Hamiltonian is written in the form:
\begin{equation}
    \hat{H}=\U + \mathbf{h} \cdot \hat{\mathbf{O}}.
\end{equation}
A bijective map between $\mathbf{h}$ and the array of expectation values $\mathbf{o}=(o_1,o_2,\dots)$, where $o_a=\braket{\hat{O}_a}$, exists~\cite{proof_hk_mapping}, provided that the ground state is not degenerate and that
\begin{equation}
    \det \left[M_{ab}\right]=\det \left[ \braket{\hat{O}_a \hat{O}_b} - \braket{\hat{O}_a} \braket{\hat{O}_b} \right] \ne 0.
    \label{determinant condition}
\end{equation}
The ground-state energy $E_{\gs}$ can be determined by minimizing a universal functional $E[\mathbf{o}]$, which satisfies the variational property:
\begin{equation}
   E_\gs  \leq E[\mathbf{o}],
\end{equation}
with the equality holding when $\mathbf{o}$ coincides with the ground-state expectation values $\mathbf{o}_{\gs}$.
The functional is conveniently written separating the universal term $F[\mathbf{o}]=u\equiv\braket{\U}$, obtaining:
\begin{equation}
    E[\mathbf{o}]= F[\mathbf{o}]+\mathbf{o} \cdot \mathbf{h}.
\end{equation}\\

It is worth stressing that the above formalism can in principle be applied to rather general Hamiltonians, including spin models, or fermionic and bosonic lattice models. The variables $\mathbf{h}$ might represent random couplings or random external fields. On the other hand, the fixed term $\U$ might represent hopping operators, external fields, or interaction terms.
Our goals are to develop accurate approximations for the unknown universal functional $F[\mathbf{o}]$ using deep learning techniques and, then, to find the ground-state properties by minimizing $E[\mathbf{o}]$ via a gradient-based optimization. For the Hamiltonians addressed in this Article, the applicability condition Eq.~\eqref{determinant condition} is always formally fulfilled. Still, one might expect that the regimes where $\det \left[M_{ab}\right]$ is finite but vanishingly small might be more susceptible to the possible residual inaccuracies of the deep-learned functional. This phenomenon is discussed in Section~\ref{subsecvarying}.

\section{Testbed Hamiltonians}
\label{section: hamiltonians}
We develop and test DL functionals addressing two  quantum spin models. The first is an integrable Hamiltonian, namely, the one-dimensional Ising model with nearest-neighbor ferromagnetic couplings and random transverse field. The second includes also next-nearest neighbor interactions, so that it is not exactly integrable. The two models will be referred to as 1nn and 2nn quantum Ising models, respectively. Further details are provided hereafter.
\subsection{Nearest-neighbor Ising model}
\label{section: 1nn model}
The one-dimensional Ising model with disordered transverse field and nearest-neighbor interactions (1nn) can be written in the form $\hat{H}=\U+\mathbf{h} \cdot \hat{\mathbf{Z}}$, where the universal term $\U=\sum_i \U_i$ can be expanded as a sum over the single-site terms:
\begin{equation}
\label{H0NN}
    \U_i=-J \X_i \X_{i+1};
\end{equation}
here, $\hat{\mathbf{Z}}=(\Z_1,\dots,\Z_l)$, $\X_i$ and $\Z_i$ are standard Pauli matrices at site $i=1,\dots,l$, $l$ is the number of spins, and $\mathbf{h}=(h_1,\dots,h_l)$ denotes the array of transverse fields. We choose a ferromagnetic coupling, namely $J>0$, also setting the energy scale used throughout this Article. The transverse fields are sampled from uniform random distributions in the range $0\le h_i \le h$.
The parameter $h\geq0$ determines the disorder strength.
Periodic boundary conditions are assumed, which corresponds to setting $\X_{l+1}\equiv \X_{1}$.
The above Hamiltonian is integrable. Its ground-state properties can be efficiently computed through the Jordan-Wigner transformation~\cite{Lieb_1961,russomanno}. This allows us to solve many Hamiltonian instances, even for large lattice sizes $l\sim 100$, at a negligible computational cost.
The ground state is known to undergo a paramagnetic to ferromagnetic phase transition when the disorder strength $h$ is reduced. For our choice of transverse fields, the critical point  can be easily computed following Refs.~\cite{PFEUTY1979_disorder,Young_1nn_ising}, obtaining $h=eJ$.

In the following, DL techniques are used to approximate the unknown universal functional $F[\mathbf{z}]=u$, where the argument is  the array $\mathbf{z}=(z_1,\dots,z_l)$ of the transverse magnetizations 
   $ z_i = \braket{\Z_i}$.
%
Notably, the additivity of the universal term $\U$ allows us implementing a scalable functional written in the form $F(\mathbf{z})= \sum_i f_{i} (\mathbf{z} )$, where $f_{i} (\mathbf{z} ) = u_i \equiv \braket{\U_i}$.
This structure allows us to train and test DL functionals on different system sizes.
It also leads one to consider the spatial properties of the functionals $f_i[\mathbf{z}]=u_i$. 
Specifically, the question is whether the output value $u_i$ is allowed depending  only on the magnetizations $z_j$, for sites $j$ in the neighborhood of the site $i$, or whether long distance couplings can be accounted for. This important issue is discussed in subsection~\ref{subsecnetwork}.
%

\subsection{Next-nearest neighbor Ising model}
The second testbed model is the one-dimensional transverse-field Ising model with up to the next-nearest neighbor interactions (2nn). The Hamiltonian can be written in the form $\hat{H}=\hat{U}+ \mathbf{h} \cdot \hat{\mathbf{Z}}$, where the universal term is $\hat{U}= \sum_i \hat{U}_{i}$, with
\begin{equation}
    \U_i=-J \left( \X_i \X_{i+1} + \X_i \X_{i+2} \right).
\end{equation}
We choose ferromagnetic couplings $J=1$, and we sample the random transverse field from a uniform distribution in the range $0\le h_i \le h$.
Periodic boundary conditions are assumed, namely, $\X_{l+1}\equiv \X_{1}$ and $\X_{l+2}\equiv \X_{2}$.
The above Hamiltonian is not integrable. For lattice size up to $l=20$, we determine the ground-state energies and the transverse-magnetization profiles  using  exact diagonalization routines~\cite{quspin}.
For larger system sizes we resort to tensor network methods~\cite{itensor}, reaching convergence in terms of bond dimension and number of optimization iterations. 
The ground state is found to undergo a ferromagnetic quantum phase transition. The critical disorder strength, averaged over many realizations of the random transverse field, is found by means of a finite-size scaling analysis of the Binder cumulant of the longitudinal magnetization~\cite{Binder1981,binder_melko}. This analysis is provided in Appendix~\ref{section: binder cumulant}. The estimated critical point is  $h = 5.60(15)J $. \\
As in Section~\ref{section: 1nn model}, our intent is to obtain the expectation values $u_i$ from a scalable functional written in the form $ F(\mathbf{z}) = \sum_i f_{i}(\mathbf{z})=\sum_i u_{i}$




\section{Density functional theory via deep learning}
\label{secdldft}
The central task of this study is to train deep neural networks to represent accurate approximations of the unknown universal functional $F[\mathbf{z}]=u$. 
From here on, the $F[\mathbf{z}]$ denotes the approximation learned by the neural networks. 
For both Hamiltonians described in Section~\ref{section: hamiltonians}, the universal part $\hat{U}$ can be written as a sum of single-site terms: $ \hat{U} = \sum_i \hat{U}_i$. Henceforth, the network can be trained to map the array of transverse magnetizations $\mathbf{z}$ to an equal length array, namely, the set of expectation values $\mathbf{f}[\mathbf{z}]=(f_1[\mathbf{z}],\dots,f_l[\mathbf{z}])=\mathbf{u}$, where $\mathbf{u}\equiv (u_1,\dots,u_l)$ .
%
Hereafter we describe the network training procedure and the gradient-based optimization used to search for the ground state. 
A schematic representation of the whole procedure is shown in Fig.~\ref{vs_training_size}.

\subsection{Network architecture and training protocol}
\label{subsecnetwork}
\begin{figure}[H]
  \centering
  \subfigure{\includegraphics[width=0.90\columnwidth]{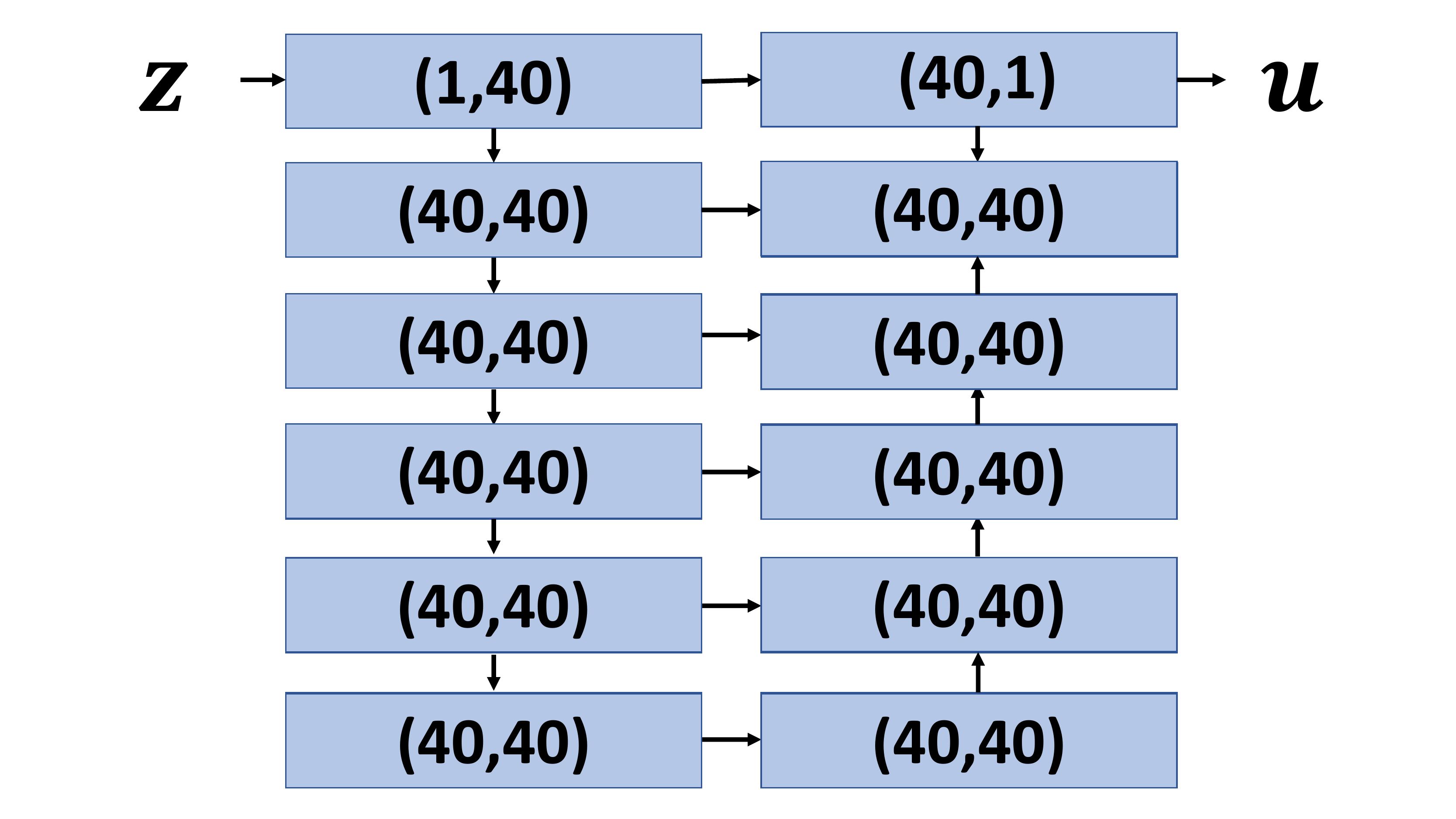} \label{unet picture}}
  \caption{Schematic representation of the U-NET neural network. The input is the transverse magnetization profile $\mathbf{z}$. This is processed through sequential convolutional layers also with skipped connections. The output $\mathrm{u}$ corresponds to the expectation values of the universal terms in the Hamiltonian. The pairs of numbers within the rectangles represent the number of input and output channels, respectively.
 }
 \label{lda_fit}
\end{figure}

Special care is taken to implement a scalable neural network, so that it can be applied to different system sizes even without re-training. Some previous studies, which  tackled quantum many-body systems within different computational frameworks, have already implemented scalability following various strategies, focusing either on the network architecture and/or on the system descriptors. These strategies include: tiling the systems via parallel networks~\cite{mills2019extensive}, fixing the output size via global pooling layers~\cite{saraceni2020scalable,Cantori_2023}, including the number of variables as an additional system descriptor~\cite{10.21468/SciPostPhys.10.3.073}, or adopting fixed-size system representations~\cite{jungsize}. Here, as in Ref.~\cite{inhomogeneous_prediction}, we exploit the linear scaling of the output length with the system size, and we implement scalability by building the network using only convolutional filters. Indeed, convolutional layers are designed to scan their input via fixed-size kernels, but the output size scales with the corresponding input size. Sequences of convolutional layers can be stacked without losing scalability.
Specifically, the input layer of our network has $l$ neurons corresponding to the $l$-component array $\mathbf{z}$ . The activations of the $l$ neurons in the output layer shall represent the target array $\mathbf{u}= \mathbf{f}[\mathbf{z}]$.
Importantly, the size of the filter kernel $k_s$, but also the number of convolutional layers $N_{\mathrm{la}}$, determine the reach of the direct spatial connections accessed by the functionals $f_i[\mathbf{z}]$. 
Specifically, the output $u_i$ my depend on the input $z_j$, as long as the modulus of the distance is $|i-j|\leq N_\mathrm{la} (k_s-1) /2$ ($k_s$ assumed odd).
%
Therefore, as soon as $k_s>1$, the DL functionals are highly non-local, and they can easily account for the possible direct coupling between input and output variables corresponding to quite distant spins, already with moderately deep networks. 
%
%
%
It is known that the type of network architecture affects the performance in simulating quantum systems~\cite{hermann2020deep,PhysRevE.101.063308,lou2023neural}. After exploring different (convolutional) network models, we choose to adopt an architecture closely inspired by the so-called U-NET~\cite{unet}. This is built via a sequence of $N_{\mathrm{la}}$ convolutional layers, allowing both standard sequential connections and also skipped connections. The network structure we adopt is visualized in Fig.~\ref{lda_fit}. Each convolutional layer includes $N_{\mathrm{hc}}=40$ channels associated to just as many filters of size $k_s=5$.
For the first $N_{\mathrm{la}}/2$ layers the hidden output at the $m-th$ layer $c_{\alpha}^{m}$  is
\begin{equation}
    c_{\alpha}^{m}=  \mathrm{Act} \left(\sum_{\beta} W_{\alpha \beta}^{m-1} c_{\beta}^{m-1} + b^{m-1}_{\alpha} \right),
\end{equation}
where $m \in [1,N_{\mathrm{la}}/2]$.
For the next $N_{\mathrm{la}}/2$ layers, the output of the $N_{\mathrm{la}}-i+1$-th layer in the channel $\alpha$, namely $c_{\alpha}^{N_{\mathrm{la}}-i+1}$, is computed as:
\begin{equation}
    c_{\alpha}^{N_{\mathrm{la}}-i+1}=  \mathrm{Act} \left(\sum_{\beta} W_{\alpha \beta}^{N_{\mathrm{la}}-i} (c_{\beta}^{N_{\mathrm{la}}-i}+c_{\beta}^{i}) + b^{N_{\mathrm{la}}-1}_{\alpha} \right),
\end{equation}
where $\mathrm{Act}$ is the activation function, $W^{n}_{\alpha\beta}$ and $b^{n}_{\alpha}$ are the weights and biases of the $n$-th layer and $i \in [1,N_{\mathrm{la}}/2]$.

For the results reported in Sections~\ref{Results}, $N_{\mathrm{la}}=12$ layers are used. A further analysis on the effect of different architectural details is reported in the Appendix~\ref{section: systematic improvement}. It is worth mentioning that  circular padding is adopted in all layers to comply with the periodic boundary condition of the physical system. 
The network is trained using datasets arranged in the format $\{\mathbf{z}^{(a)},\mathbf{u}^{(a)}\}$, where the superscript $(a)$ labels different random realizations of the Hamiltonian.
The loss function is the mean squared error 
$\mathrm{MSE}= \frac{1}{N_{\mathrm{tr}}} \sum_{a=1}^{N_{\mathrm{tr}}} \left| \mathbf{u}^{(a)}-\mathbf{f}[\mathbf{z}^{(a)}] \right|^2$, where $N_{\mathrm{tr}}$  denotes the number of instances in the training dataset.

%
%
The weights and biases are optimized using the ADAM algorithm~\cite{kingma2014adam} with batch size $bs=100$, training learning rate $lr=10^{-4}$, iterated for 3000 epochs.
The training datasets include $N_{\tr} \in [5000,1.5\times10^5]$ instances, for system sizes in the range $l \in [8,24]$. The test datasets include $N_{\ts} \in [200,6000]$ instances, for system sizes $l \in [8,64]$.
The random fields $h_i$ of the training and testing Hamiltonians are sampled by a uniform distribution in the interval $[0,h]$ and we consider the critical disorder strength $h=eJ$, as well as two values in the paramagnetic and ferromagnetic phases, namely, $h=3.5J$ and $h=1.8J$, respectively.

\subsection{Gradient-descent minimization}
Once the network is trained, the ground-state magnetization profile corresponding to previously unseen  Hamiltonian instances can be searched for by minimizing the DL prediction $e_{\mathrm{DL}}(\mathbf{z})$ for the energy per spin, which reads:
\begin{equation}
    e_{\mathrm{DL}}(\mathbf{z})=\frac{1}{l}\left[\sum_{i=1}^{l}  f_i[\mathbf{z}] + \mathbf{h} \cdot \mathbf{z} \right].
\end{equation}
The exact ground-state energy per spin $e_{\gs}=E_{\gs}/l$ can then be estimated from the prediction corresponding to the optimal profile $\mathbf{z}_{\mathrm{min}}$, namely $e_{\mathrm{min}} =e_{\mathrm{DL}}(\mathbf{z}_{\mathrm{min}})$.
%
%
The minimization is performed by iterating gradient-descent updates. To take into account the constraint $z_i \in [-1,1]$, a change of variables is performed introducing the angles $\theta_{i}$, such that
    $z_{i}=\cos(\theta_{i})$ for $i=1,\dots,l$.
Thus, the update at step $t=0,1,\dots$ is performed as:
\begin{equation}
    \theta_{i, t+1} =\theta_{i, t} - \eta \frac{\partial e_{\mathrm{DL}}(\mathbf{z}_t)}{\partial \theta_{i,t}},
\end{equation}
where $\eta>0$ is a small learning rate, while $\theta_{i, t}$ and $\mathbf{z}_t$ denote the angles and magnetization profile at step $t$, respectively.
A schedule with a fixed learning rate $\eta=0.1$ is found to be suitable for all system sizes. Different maximum numbers of gradient-descent steps are considered in the range  $t \in [2000,5000]$, depending on the size of the system, with larger sizes requiring more steps to converge to the minimum configuration.
The minimization appears  not to be affected by local minima; indeed, runs started from different initial random angles $\theta_{i,0}$ converge to the same state.
It is also worth pointing out that the gradient-descent procedure turns out not to affected by the instabilities encountered in the use of DL-functionals in continuous-space systems~\cite{snyder_rupp_machine_learning,PhysRevLett.108.253002,https://doi.org/10.1002/qua.25040,Meyer2020,costa_scriva}. In that setup, even small errors in the functional derivatives can create  nonphysical density profiles, which significantly differ from the ground-state profiles used to train the DL functionals. In that case, the network might provide erroneous predictions that lead to energies even well below the ground state, henceforth violating the variational property held by the exact functional. 
Several remedial strategies have been adopted, including, e.g, gradient projection~\cite{Meyer2020,snyder_rupp_machine_learning} or interchannel averaging layers~\cite{costa_scriva}. As shown in Section~\ref{Results}, the DL functional implemented here for quantum Ising models do not violate the variational property, in the interesting physical regime, beyond controllable statistical fluctuations.

In Section~\ref{Results}, the following figures of merit are used to quantify the performance of our DL-DFT method. The first is the relative energy error $\Delta_r e \equiv ( e_{\gs}-e_{\mm}) / e_{\gs}$ of the gradient-descent prediction $e_{\mm}$ from the corresponding exact ground-state results $e_{\gs}$. Its disorder average $\overline{\Delta_r e}$ is computed as
\begin{equation}
  \overline{\Delta_r e} = \frac{1}{N_{\ts}} \sum^{N_{\ts}}_{a=1}\frac{e^{(a)}_{\gs}-e^{(a)}_{\mm} }{e^{(a)}_{\gs}} ,
\end{equation}
where $N_{\ts}$ is the number of test Hamiltonian instances. The second is the analogous relative energy error   in absolute value $|\Delta_r e|$.
%
The next figure of merit, $\Delta_r z$, quantifies the accuracy of the magnetization profiles provided by gradient descent. Its disorder average is:
\begin{equation}
     \overline{\Delta_r z } = \frac{1}{N_{ts}} \sum^{N_{\ts}}_{a =1}\frac{\left|\left|\mathbf{z}^{(a)}_{\gs}-\mathbf{z}^{(a)}_{\mm}\right|\right|}{||\mathbf{z}^{(a)}_{\gs}||},
\end{equation}
where $\left|\left| . \right|\right|$ represents the $L_1$ norm.
We also consider the predictions provided by the trained functional when fed with exact ground-state magnetization profiles $\mathbf{z}_{\gs}$. Specifically, we focus on the universal contribution 
$u$, inspecting the relative error compared to the exact ground-state result, namely, 
$\Delta_r u=\left[
\sum_{i=1}^l f_i
[\mathbf{z}_{\gs}] -u\right]/u
$.
Notice that this metric only accounts for the inaccuracy of the trained functional, thus excluding possible inaccuracies in $\mathbf{z}_{\mm}$ due to  errors occurring in the gradient-descent optimization.
It is worth mentioning here that, in Section~\ref{subsec: scaling law analysis}, also the accuracy of the predictions for the input-output statistical correlations is analyzed, but at a qualitative level.

\begin{figure}[h!]
    \centering
    \subfigure{
        \includegraphics[width=0.85\columnwidth]{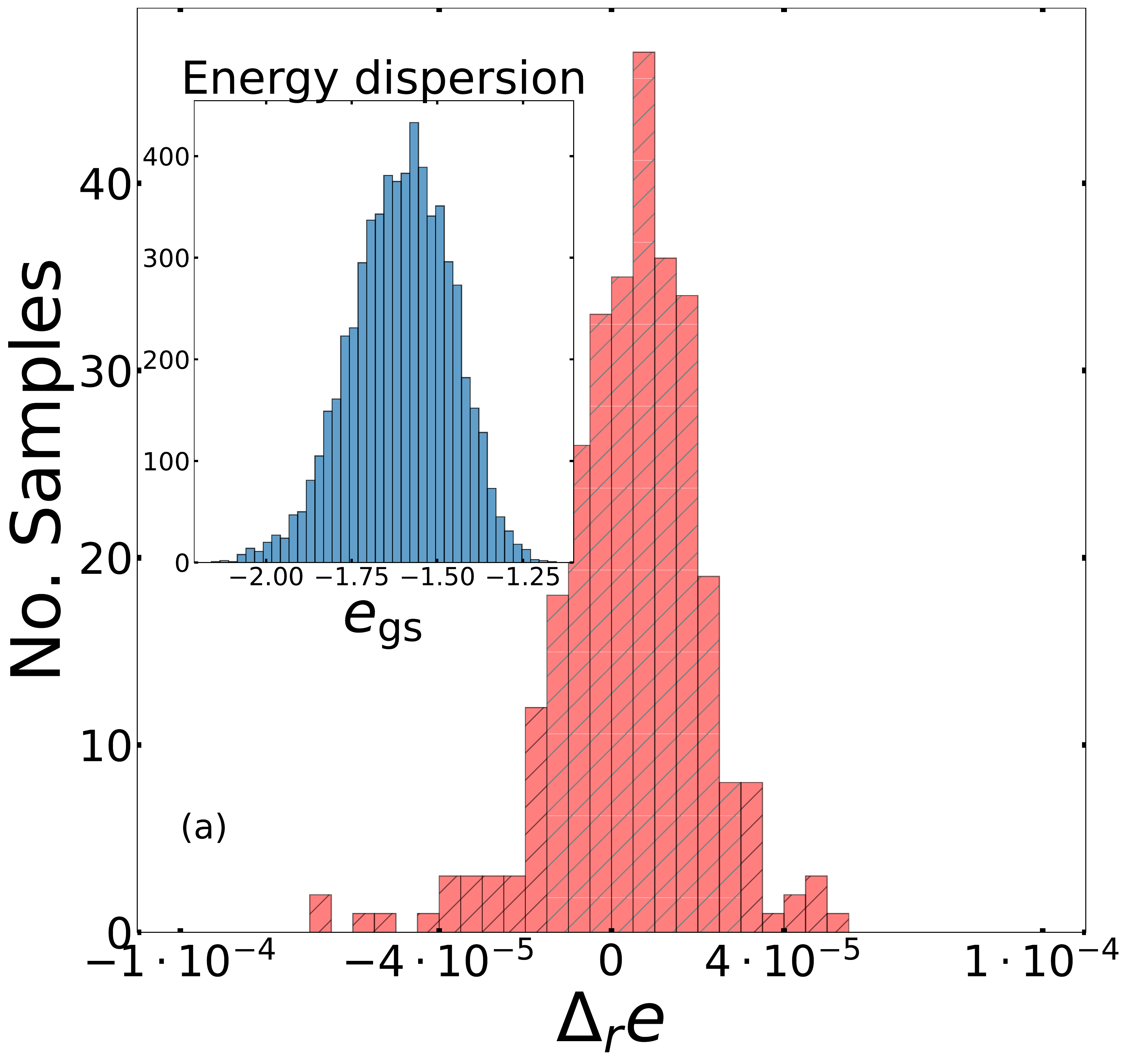}
        \label{hist_1nn eng}
    }
    \subfigure{
        \includegraphics[width=0.85\columnwidth]{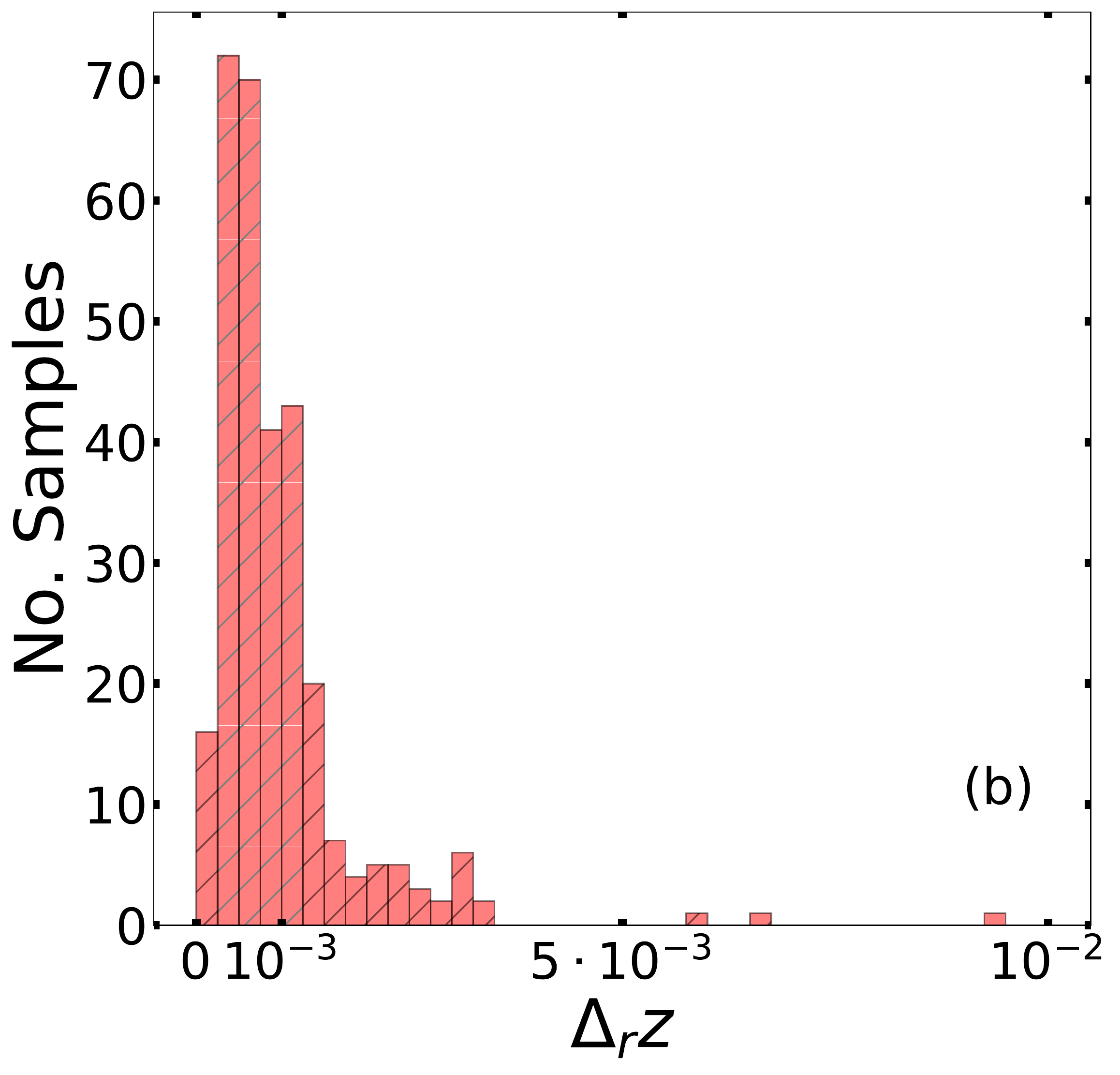}
        \label{hist_1nn z}
    }    
     \caption{
Histograms of the relative energy discrepancies $\Delta_r e $ [panel (a)] and the magnetization discrepancies [panel(b)] $ \Delta_r z $ for the $1nn$ Ising model at the ferromagnetic critical point. The size of the system is $l=16$.  
$N_{\mathrm{ts}}=300$ disorder instances are considered, and $t_{\mathrm{max}}=2000$ steps of the gradient-descent optimization are performed. The inset in panel (a) shows the distribution of the ground state energies for $N_{\mathrm{ts}}=6000$ realizations of the $1nn$ Ising model at the critical disorder strength.
 }
\label{hist_16_size_1nn}
\end{figure}

\begin{figure}[h!]
    \centering
    \subfigure{
        \includegraphics[width=0.85\columnwidth]{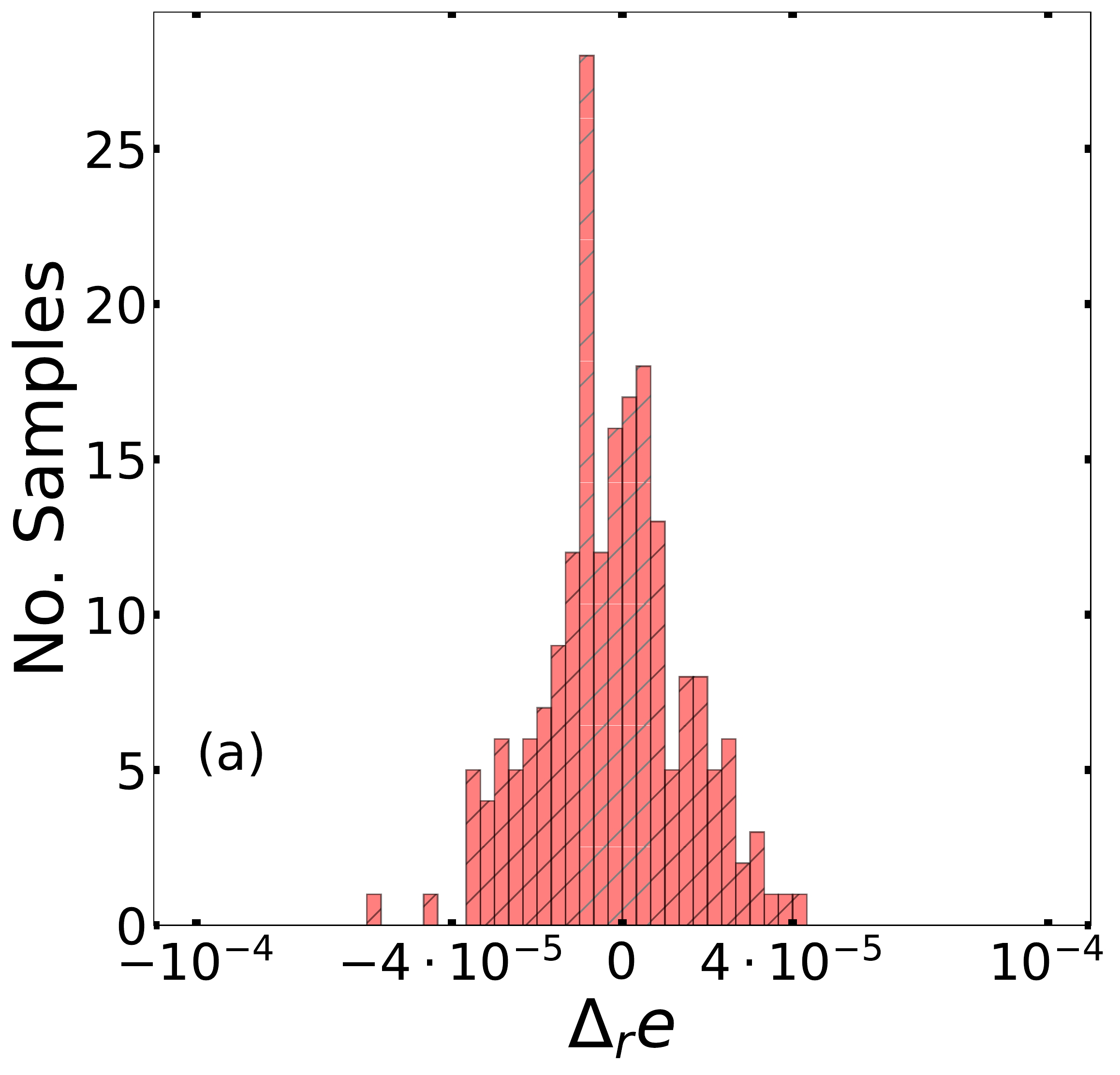}
        \label{hist_2nn eng}
    }
    \subfigure{
        \includegraphics[width=0.85\columnwidth]{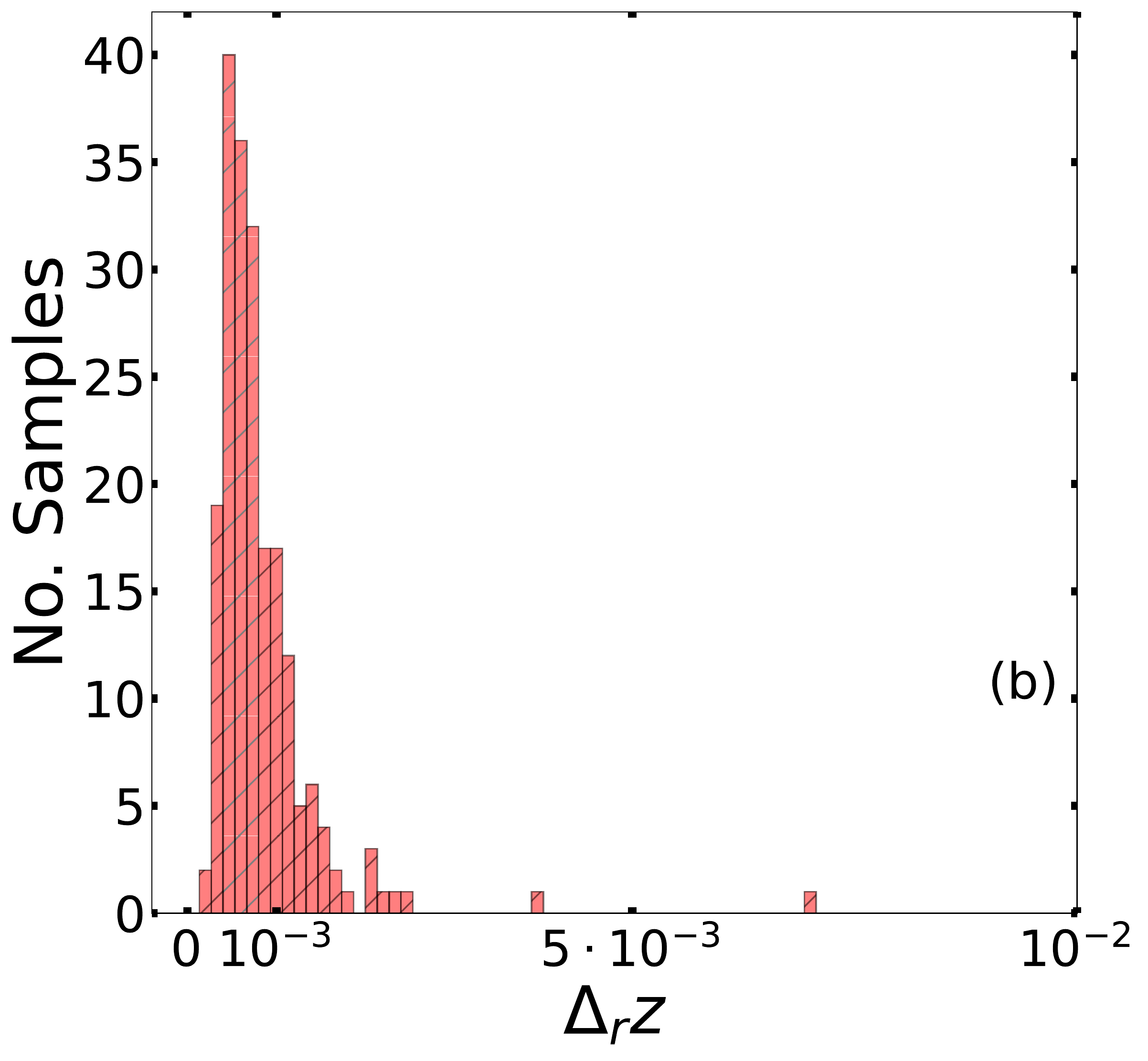}
        \label{hist_2nn z}
    }
    
    \caption{
Histograms of the relative energy discrepancies $\Delta_r e $ [panel (a)] and the magnetization discrepancies [panel(b)] $\Delta_r z$ for the $2nn$ Ising model at the ferromagnetic critical point. The size of the system is $l=16$.
$N_{\mathrm{ts}}=300$ disorder instances are considered, and $t_{\mathrm{max}}=2000$ steps of the gradient-descent optimization are performed.}
    \label{hist_16_size_2nn}
\end{figure}

\section{Results for gradient-descent optimization}
\label{Results}
\subsection{Accuracy at the quantum critical point}
The accuracy of our DL-DFT approach is tested addressing the two random Ising models described in Section~\ref{section: hamiltonians}. In this subsection, the gradient descent is performed on the same chain length used for training the network, which is set here at $l=16$. 
We first report the performance metrics computed at the quantum critical point, 
since it is expected to correspond to the worst-case scenario.
This issue is further elaborated on in Section~\ref{subsec: scaling law analysis}. 
The relative energy discrepancies $\Delta_r e$ and the ground-state energy distribution for the 1nn Hamiltonian are shown in Fig~(\ref{hist_1nn eng}). 
The average relative error in absolute value
is as small as $\overline{|\Delta_r e|}  =1(1) \cdot 10^{-5}$ (the number in parentheses indicates the standard deviation). 
This is consistent with the residual prediction error $\Delta_r u = 2.9(4) \cdot 10^{-5}$ computed on exact magnetization profiles. 
One also notices that negative errors are approximately as likely as the positive ones. This implies that no statistically significant violation of the variational property occurs. 
The accuracy of the (transverse) magnetization predictions is analyzed in Fig.~(\ref{hist_1nn z}). The average relative error is $\overline{ |\Delta_r z|} = 0.001(1) $.
%
%
A similar accuracy is reached for the 2nn Ising models (see Figs.~\ref{hist_2nn eng} and \ref{hist_2nn z}). Specifically, we obtain $\overline{|\Delta_r e  |}  = 1(1) \cdot 10^{-5}$ and  $\overline{\Delta_r z }= 0.0008(6)$ . 
These performances have been obtained using networks featuring $N_{\mathrm{la}}=12$ layers, trained on $N_{\mathrm{tr}}=15 \cdot 10^4$ Hamiltonian instances. It is worth stressing that, in fact, the performance can be systematically improved by increasing the network depth and the training dataset size. This is shown in Appendix~\ref{section: systematic improvement}.

\subsection{Accuracy for varying disorder strength}
\label{subsecvarying}
\begin{figure}[h!]
  \centering
  \subfigure{
  \includegraphics[width=0.90\columnwidth]{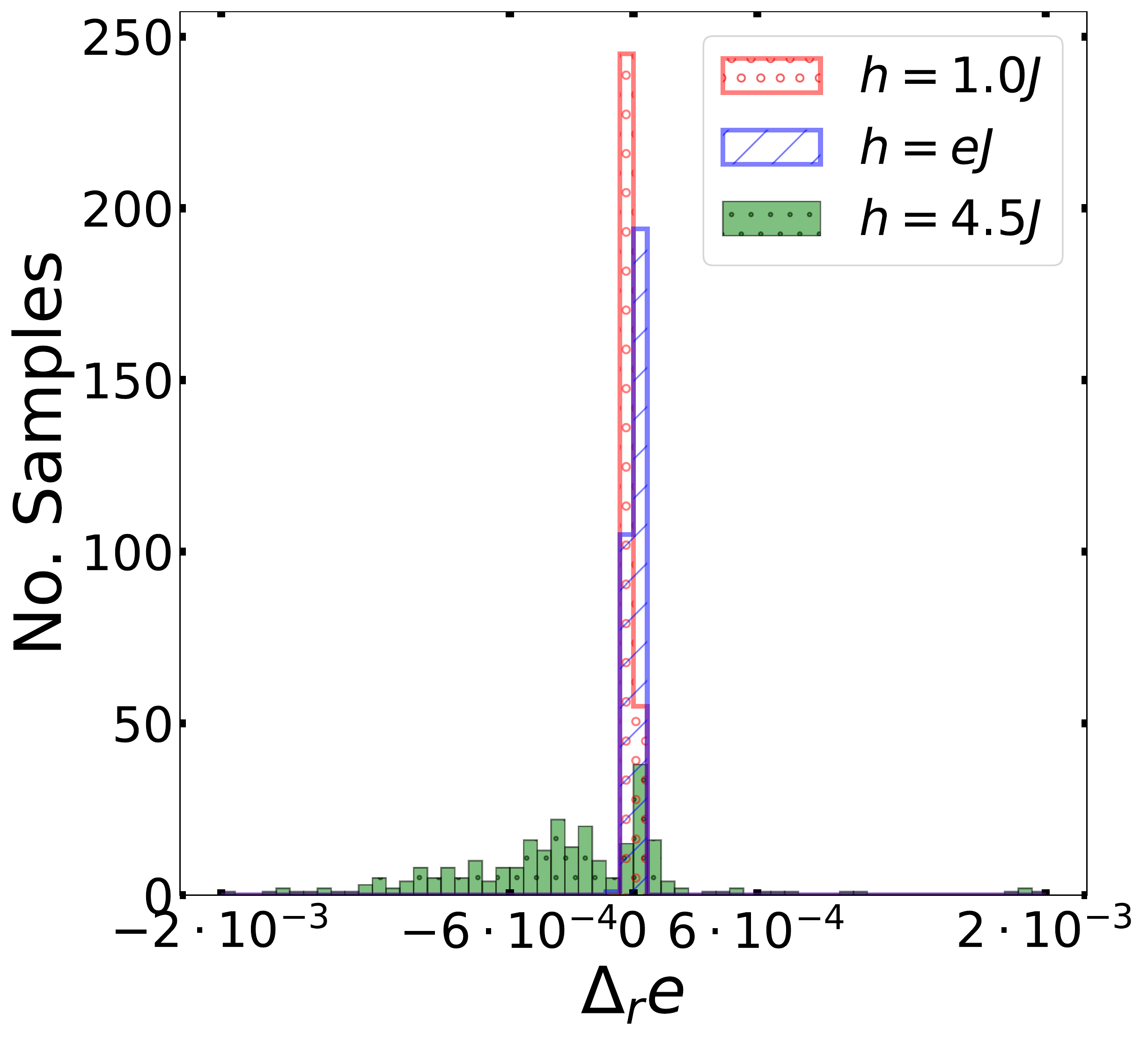} \label{e_hist_lda}}
  \caption{
Histograms of the relative energy discrepancies $\Delta_r e $ for different disorder strengths $h$. The test set includes $N_{\ts}=300$ instances of the $1nn$ Ising Hamiltonian for a chain length $l=16$. 
 }
 \label{vs disorder}
\end{figure}

\begin{figure}[h!]
\centering
\includegraphics[width=0.90\columnwidth]{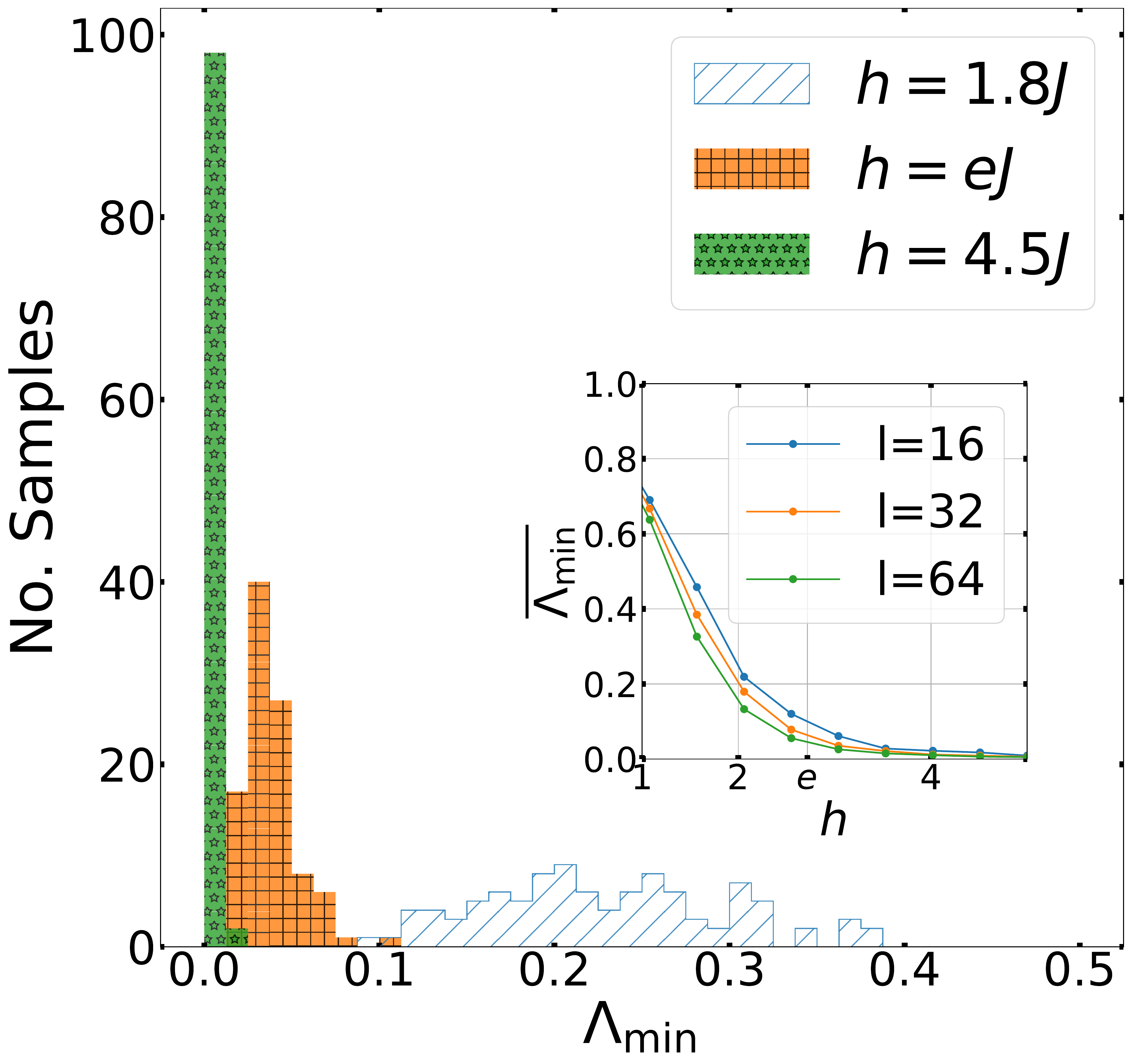}
\caption{Main panel: Histogram of the minimum eigenvalue $\Lambda_{\mathrm{min}}$ of the matrix with entries $M_{ij}=\braket{\Z_i \Z_J} -\braket{\Z_i} \braket{\Z_j}$. $N_{\ts}=100$  instances of the $1nn$ Ising Hamiltonian are considered, for three disorder strengths $h$. Inset: Average $\overline{\Lambda_{\mathrm{min}}}$ over 200 disorder realizations of the minimum eigenvalues as a function of the disorder strength $h$, for three chain lengths $l$.}
\label{detM}
\end{figure}

%
Here we analyze the  accuracy of our DL-DFT method for different disorder strengths. The histograms shown in Fig.~\ref{vs disorder} represent the energy errors after gradient descent for different instances of the 1nn Ising model, at the critical point as well as into the paramagnetic and the ferromagnetic phases. One observes an accuracy degradation at strong disorder, leading also to small violations of the variational property for some Hamiltonian instances. These are due to instabilities in the gradient-descent process~\footnote{At large $h$, we also exploit a data augmentation method to partially suppress these instabilities. Each instance in the training dataset is duplicated flipping the sign of $\mathbf{z}$, since the target $\mathbf{u}$ remains unchanged.}. 
Instead, instabilities do not occur at the critical point and in the ferromagnetic phase.
To rationalize the degradation in the paramagnetic phase, we consider the conditions required to ensure a bijective map from $\mathbf{h}$ to $\mathbf{z}$ (see Section~\ref{sectionmethods}). At strong disorder $h\gg 1$, some components of the transverse magnetization might saturate close to $z_i\simeq -1$. 
Therefore, changes in $h_i$ might determine numerically imperceptible changes in $z_i$, essentially leading to an effective breakdown of the bijective property. To quantify this effect, we analyze to what extent the condition defined by Eq.~\eqref{determinant condition}, which ensures the bijective property, is fulfilled. Following Ref.~\cite{proof_hk_mapping}, we compute the smallest eigenvalue of the matrix $\left[M_{ij}\right]$ (see Eq.~\eqref{determinant condition}), since this allows detecting when the determinant vanishes and, therefore, when the bijective property is not granted. As shown in Fig.~\ref{detM}, vanishingly small eigenvalues occur for large $h$, which is consistent with the accuracy degradation discussed above.
While this denotes a possible limitation of the DL-DFT method, it is worth pointing out that the degradation occurs only for very large transverse fields, away from the most relevant regime in the vicinity of the critical point. For large transverse field values, alternative methods, e.g., perturbation theory, are preferable (see, e.g., Refs.~\cite{PhysRev.100.36,PhysRevE.92.022118,PhysRevB.94.125109} and Refs.~\cite{PhysRevB.63.224401,chakrabarti2008quantum,Mossi_2017,PhysRevB.81.064412} for applications to Ising models). Furthermore, an alternative formulation of the DL-DFT, namely, one based on the longitudinal magnetization, would be more suitable.

\subsection{Comparison with the local density approximation}
In the case of continuous-space systems, the most popular approaches to build the universal functional start from the so-called local density approximation (LDA), eventually including semi-local correction terms. The LDA has been adapted to quantum Ising models~\cite{dft_ising_chain}, also considering a nearest-neighbor correction. Within the LDA, the universal functional $F[\mathbf{z}]=\sum_i f_i[\mathbf{z}]$ is computed using the approximation
    $f_{\mathrm{LDA},i}[\mathbf{z}] = f_{\mathrm{homo}}(z_i)$,
where $f_{\mathrm{homo}}(z)=u_{\mathrm{homo}}/l$ denotes the ground-state energy of a homogeneous Ising chain, i.e., one featuring constant transverse fields $h_i=h$ for $i=1,\dots,l$ and, henceforth, uniform magnetization.

\begin{figure}[h!]
  \centering
  \subfigure{\includegraphics[width=0.90\columnwidth]{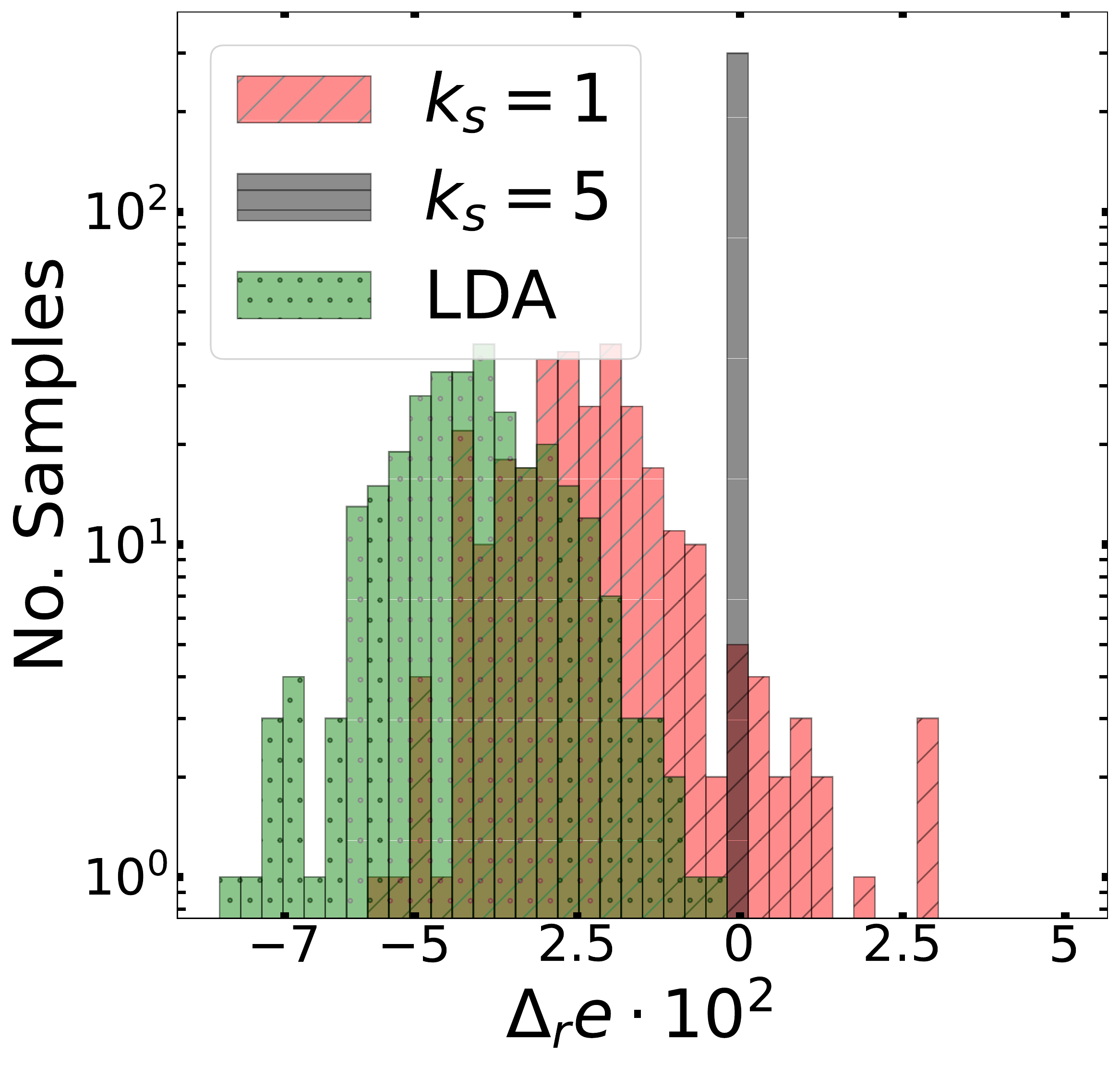} }
 \caption{
Histograms of the relative energy discrepancies $\Delta_r e$ for two networks with different kernel sizes $k_s$, and for the local density approximation (LDA).  The test includes 300  instances on the $1nn$ Ising Hamiltonian at $h=eJ$.
The training and test chain lengths are $l=16$.
 }
 \label{lda comparison}
\end{figure}

\begin{figure}[h!]
  \centering
  \subfigure{\includegraphics[width=0.85\columnwidth]{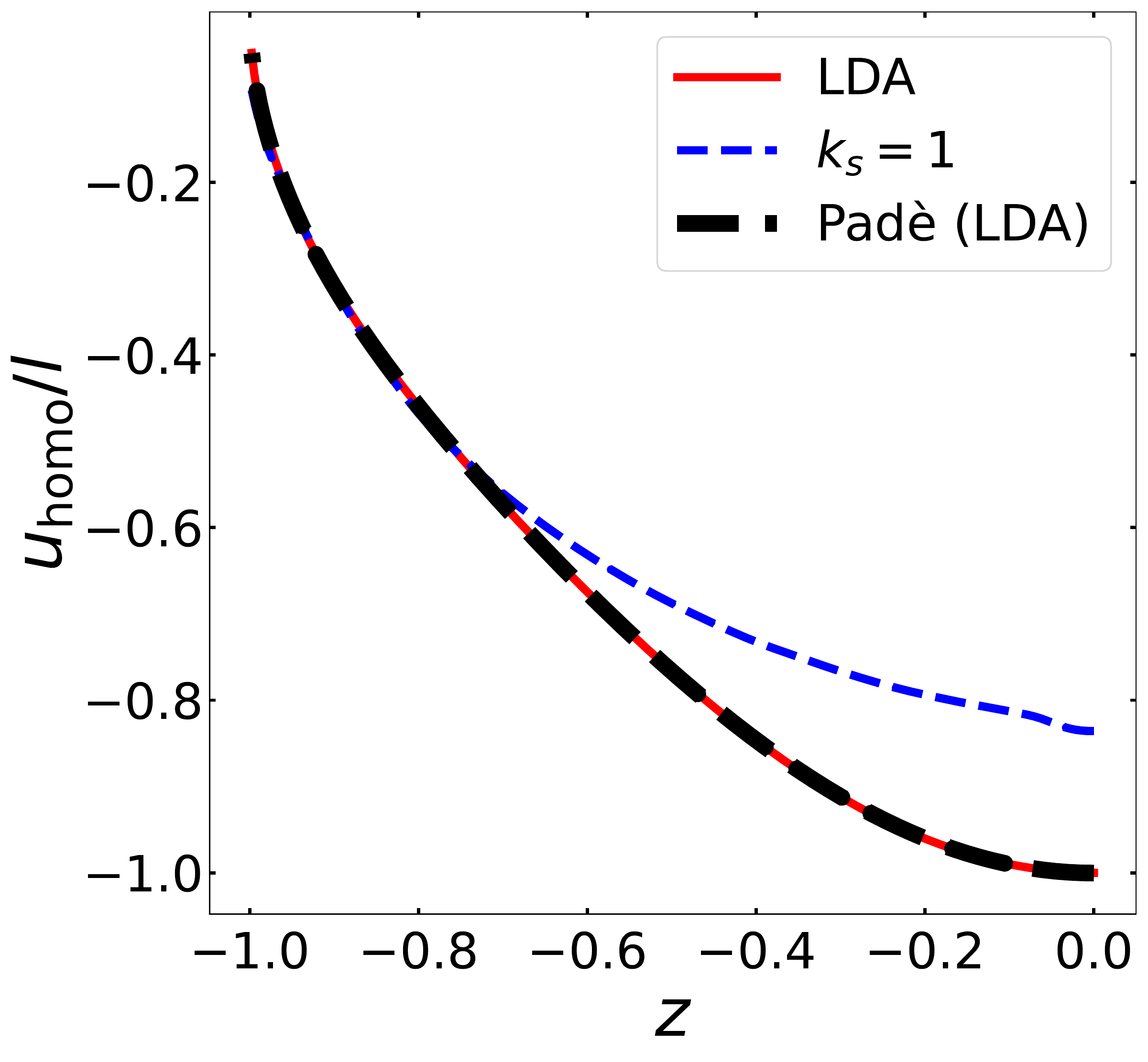} }
 \caption{Predicted value of $u_{\mathrm{homo}}/l$ as function of the input magnetization $z$ for homogeneous $1nn$ Ising models. The LDA is compared with the local neural network with kernel size $k_s=1$. The Pad\`e fit on the LDA data is also shown.} 
 \label{lda u vs z}
\end{figure}

It is interesting to compare the performance of our DL functionals against the plain vanilla LDA. We consider the 1nn Ising chain at the critical point. The LDA functional is determined by performing a best fit on the Jordan-Wigner results for the universal term $u/l$ as a function of the transverse magnetization $z$. The fitting function is a Pad\`e approximant of order $[3/4]$.
%
%
Notice that a purely local functional can be obtained also using our scalable neural networks, simply adopting kernels of size $k_s=1$. 
The comparison is shown in Fig.~\ref{lda comparison} for the chain length $l=16$. 
Notably, even the $k_s=1$ DL functional is more accurate than the LDA. To rationalize this, we show in Fig.~\ref{lda u vs z} the local function $f_{k_s=1}(z)$ corresponding to the network with $k_s=1$, making comparison with the homogeneous approximation $f_{\mathrm{homo}}(z)$. One notices good agreement for $z \lesssim 0.7$, while the two curves substantially differ for larger values of $z$. 
This implies that, for discrete systems, assuming a homogeneous system featuring a uniform magnetization equal to the local value does not lead to the optimal purely local DFT. In contrast, for continuous-space electronic systems, the equation of state of the homogeneous electron gas is considered to be at least close to the optimal choice~\cite{perspective_dft}. We attribute this discrepancy to the absence of spatial correlations in the random fields, as compared to the external fields $v(x)$, which enter the continuous-space DFT; indeed, these latter functions usually display smooth spatial profiles.
Furthermore, in the case of electronic systems, the Hartree term, which accounts for the mean-field direct interactions, is commonly separated from the universal functional.
Incidentally, as expected all curves turn out to be essentially flat for $z\simeq 0$, consistently with the symmetry under sign change of the argument $z$.
However, we observe that away from $z=0$ our DL functionals do not automatically fulfill this symmetry, even for $k_s>1$. This issue can be overcome by performing data augmentation in the training dataset, i.e., randomly adding some configurations with signed-reversed magnetization. In this case, the symmetry is accurately reproduced. For example, for the $1nn$ Hamiltonian at the critical disorder strength  at size $l=16$, the relative prediction error related to the mirror symmetry is as small as $\overline{| \Delta_r u|}=\frac{1}{N_{\mathrm{ts}}}\sum_{a=1}^{N_{\mathrm{ts}}} \frac{\sum_i^l|{f}_i[\mathbf{z}^{(a)}]-{f}_i[-\mathbf{z}^{(a)}]|}{|\sum_{i=1}^l {f}_i[\mathbf{z}^{(a)}]|}= 8(3) \cdot 10^{-5}$.
Importantly, the DL-functional with $k_s=5$ drastically outperforms the LDA (see Fig.~\ref{lda comparison}).
It is worth iterating that, as soon as $k_s>1$, the functionals $f_i[\mathbf{z}]$ based on deep networks become highly non-local since the spatial extent of the allowed couplings increase layer by layer (see discussion in sub-section~\ref{subsecnetwork}).

To conclude the discussion on local functionals, it is worth further emphasizing that, in Ref.~\cite{dft_ising_chain}, non-local corrections to the LDA have been considered. A functional inspired by  so-called generalized gradient approximation (GGA) was implemented to take into account nearest-neighbor effects. This led to a significant improvement in the maximum error, from  an error of $10\%$ with the LDA to $2.5 \%$ using the GGA.

\begin{figure}[h!]
  \centering
  \subfigure{\includegraphics[width=0.90\columnwidth]{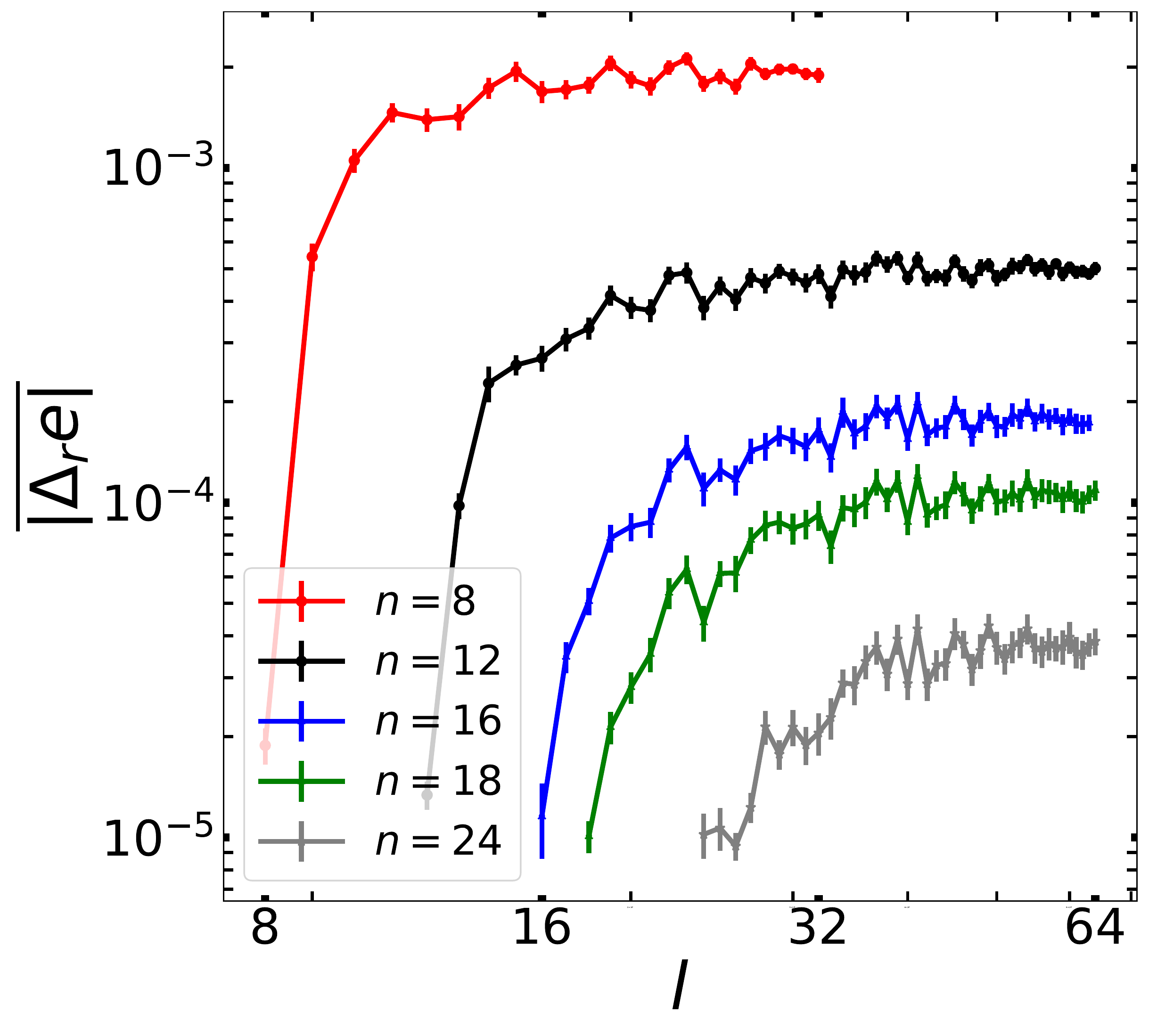}}
 \caption{
Average energy discrepancy in absolute value $\overline{|\Delta_r e|}$  as a function of the test chain length $l$. The five datasets correspond to different training chain lengths $n$. This analysis is performed on the $1nn$ Ising Hamiltonian at the critical point $h=eJ$.}
 \label{scaling gd error}
\end{figure}

\subsection{DL-DFT for predicting energies at larger sizes}
\label{subsec: scaling law analysis}
The scalable architecture of our neural networks allows applying the DL functionals to lattices of different lengths. In particular, a functional trained on small lattices can be used to predict the properties of larger systems. However, if the training size is too small, the DL functional might learn spurious finite-size effects, leading to systematic biases in the predictions on the larger test sizes $l$. This effect is investigated hereafter.
Notice that, in this subsection, we indicate the training lattice size with $n$, while for the testing size we keep the symbol $l$.
In Fig.~\ref{scaling gd error}, the average relative error of the gradient descent procedure $\overline{|\Delta_r e|}$ is analyzed, considering the 1nn Hamiltonian at the critical point. One notices that, after an initial increase, this error metric saturates as the testing systems size increases beyond $l \gtrsim 2 n$. Importantly, the asymptotic error systematically decreases as a function of $n$.

\begin{figure}[H]
  \centering
  \subfigure{\includegraphics[width=0.90\columnwidth]{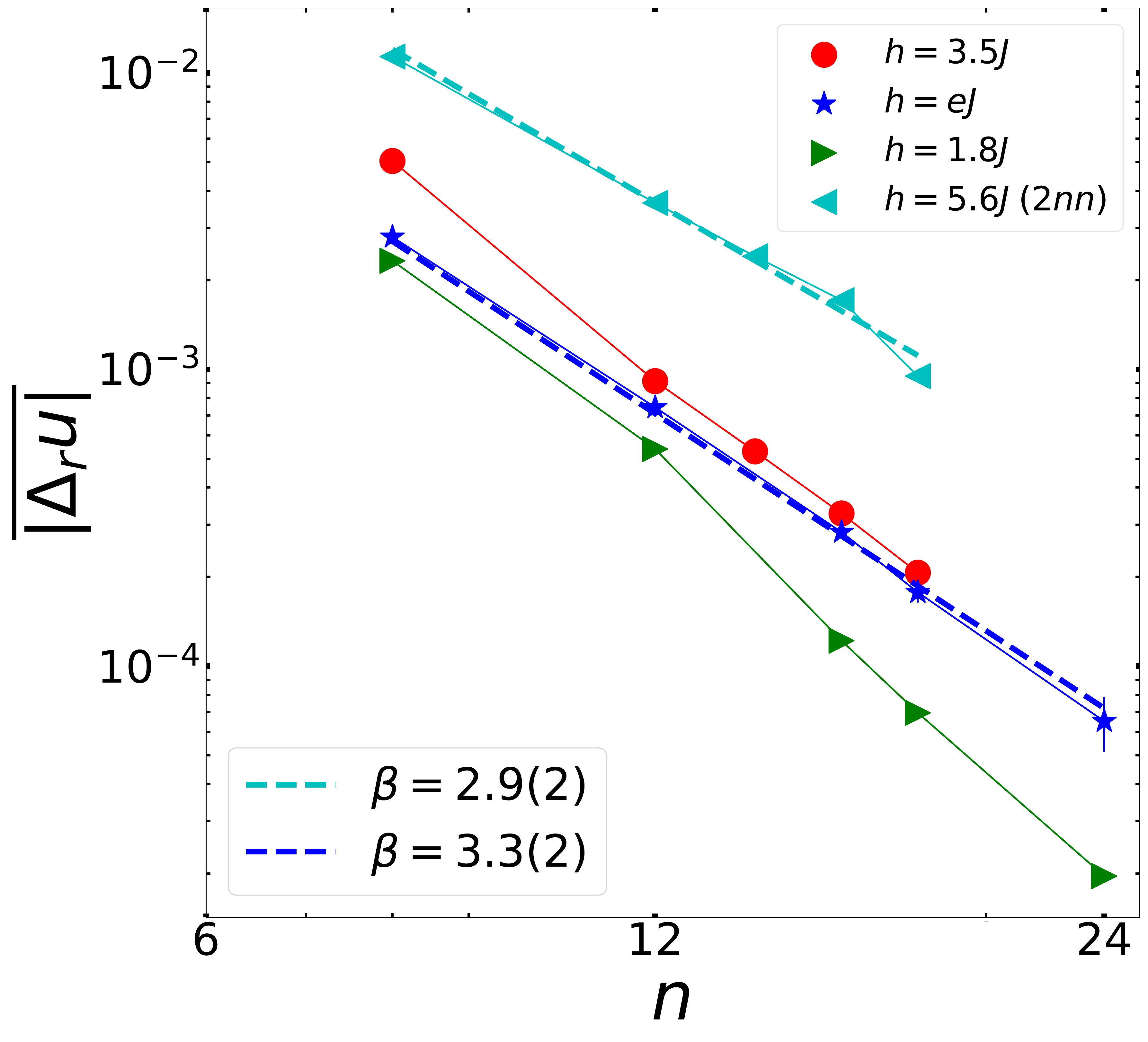} }
 \caption{Prediction error $\overline{|\Delta_r u|}$ as a function of the training chain length $n$. The testing length is $l=64$. A set of $N_{\mathrm{ts}}=1000$ samples is considered. The four datasets correspond to three  disorder strengths of the 1nn Ising model, including the critical point, and to the 2nn Ising model at its estimated critical point.
 The dashed lines represent power-low fits with negative exponent $\beta$. 
 Continuous connecting segments are guides to the eye.
 }
 \label{scaling plots:1}
\end{figure}

The scaling with $n$ of the asymptotic average error is further analyzed in Fig.~(\ref{scaling plots:1}). Only the average prediction accuracy $\overline{|\Delta_r u|}$ (computed on exact magnetization profiles) is considered, since the gradient descent error $\overline{|\Delta_r e|}$ generally displays an analogous behavior.
%
The testing system size is $l=64$, which is suitable to represent the asymptotic large-size regime, for all training sizes $n$ we consider. 
Interestingly, for the system sizes $n$ we address, at the critical point the scaling is well described by a power-law fitting function in the form
\begin{equation}
    \overline{|\Delta_r u|} \propto \frac{1}{n^{\beta}},
\end{equation}
with the exponent $\beta$ obtained from a best-fit analysis.
The 1nn and the 2nn Hamiltonians provide comparable exponents $\beta \simeq 3$.
While this might suggest a possible universal behavior, further models should be analyzed to corroborate this supposition.
Notice that,  away from the critical points, the error decays even faster, possibly more rapidly than a power-law behavior, but the accessible system sizes do not allow identifying the scaling law.
Notice that, as discussed in subsection~\ref{subsecvarying}, when the DL-DFT functional is tested on the size $l=n$, the inaccuracy amplitude is larger at strong disorder $h\gg 1$, deep into the paramagnetic phase. Here, we find that, when the test is perform on a size $l\gg n$, the scaling with $n$ of the inaccuracy is the slowest at the critical point.

\begin{figure}[H]
  \centering
  \subfigure{\includegraphics[width=0.9\columnwidth]{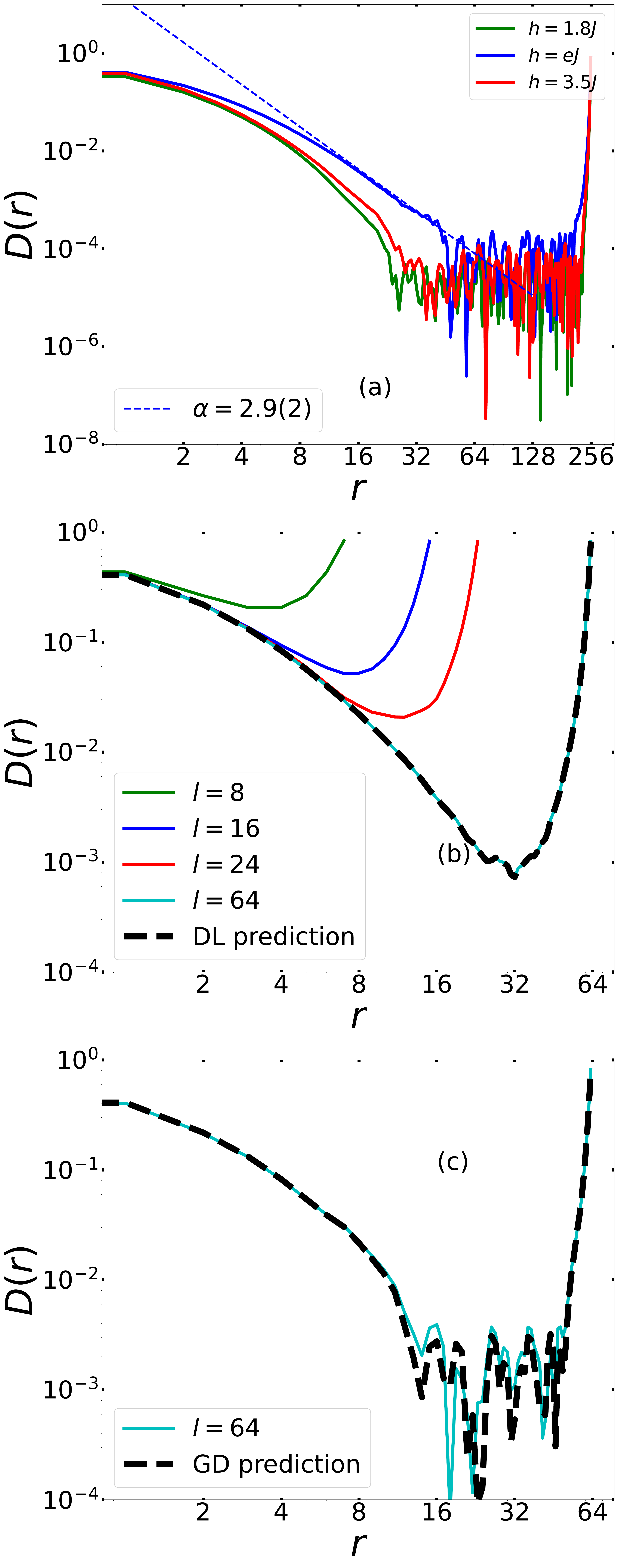} \label{covariance input output}}
 \caption{Panel (a): Normalized covariance $D(r)$ (see Eq.~\eqref{eqdr}) as a function of the distance $r$. The three datasets correspond to three disorder strengths $h$ of the $1nn$ Ising model, for a system size $l=256$. The dotted line represents a power-low fit at the critical point $h=eJ$ in the range $r \in [12,32]$ with exponent $\alpha$. 
 Panel (b):  Normalized covariance $D(r)$ for different system sizes $l$ at the critical point $h=eJ$. The dashed (black) curve represents the covariance computed with the output of the DL functional trained in the system size $n=16$, and tested on the size $l=64$. 
 Panel (c): Comparison between the normalized covariance $D(r)$  obtained performing gradient descent (GD) optimization using a DL functional trained on the system size $n=16$, and the one computed using exact expectation values, at the testing size $l=64$. In this panel, the number of samples is $N_{\ts}=6000$, for the critical point $h=eJ$ of the 1nn Ising Hamiltonian.
 }
 \label{scaling plots:2}
\end{figure}

To shed further light on the role of the lattice size on the DL-DFT approach, we investigate the statistical correlations, computed over the disorder realizations, among the inputs $\mathbf{z}$ and the outputs $\mathbf{u}$. Specifically, we compute the (translational invariant) normalized covariance, defined as:
\begin{equation}
\label{eqdr}
    D(r)=\frac{1}{l}\sum^{l}_{i=1} C_{i, i+r},
\end{equation}
where
\begin{equation}
    C_{i, j}=\frac{ \overline{z_i u_j} - \overline{z_i} \; \overline{u_j} }{\sigma(z_i) \sigma(u_j)},
\end{equation}
with $\sigma()$ the standard deviation over the disorder realizations. 
It is worth pointing out that the function $D(r)$ does not describe quantum correlations in the ground-state of a specific Hamiltonian, but rather statistical correlations of expectation values in the ensemble of Hamiltonian instances.
Here, $D(r)$ is computed considering exact ground-state expectation values.
Notably, as shown in the panel (a) of Fig.~\ref{scaling plots:2}, at the critical point and for sufficiently large lattices, the long-distance decay of the correlation function $D(r)$ appears to be well described by the power-law  
   $ D(r) \propto \frac{1}{r^{\alpha}}$,
and a best-fit analysis in the region $r\in [12,32]$ for $l=256$ provides the exponent $\alpha\simeq 3$. The decay is faster away from criticality.
These observations suggest a possible connection with the decay of the prediction error $\overline{|\Delta_r u|}$. This is further elaborate below.
One notices that the long-distance correlations are strongly affected by finite size effects [see panel (b) of Fig.~\ref{scaling plots:2}].
This might suggest that predictions of DL-functionals trained on small lattices would display strongly biased input-output correlations when tested on larger sizes. Notably, we find this is not the case. This is first demonstrated in panel (b), for the case in which the correlation function $D(r)$ is computed using the outputs $u_j$ predicted by networks fed with exact ground-state magnetization profiles. Second, in panel (c), we show the correlations among  input and output expectation values obtained from the gradient-descent search of the ground states. Notice that in the latter case only $N_{\ts}=6000$ are used, leading to sizable statistical fluctuations compared to the exact ensemble average. Anyway, in both cases the exact correlation function $D(r)$ is closely reproduced.
In conclusion of this analysis, it is worth pointing out that the statistical correlations discussed here do not imply a direct functional dependence between inputs $z_j$ and (possibly distant) output $u_i$. Unfortunately, the latter dependence cannot be easily inspected since the ground-state expectation values cannot be independently tuned at will.

\begin{figure}[H]
  \centering
  \subfigure{\includegraphics[width=0.9\columnwidth]{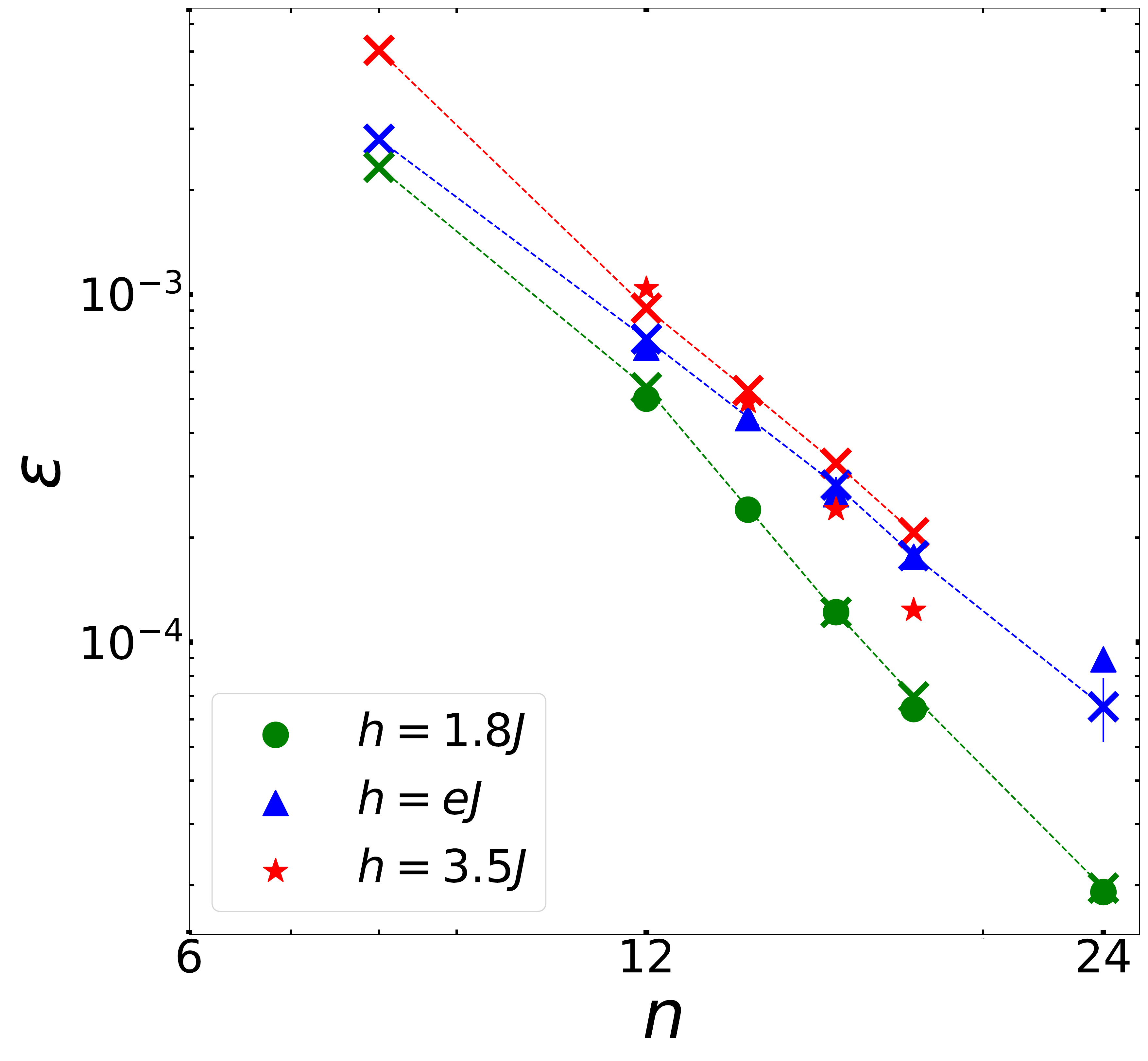}}
 \caption{Rescaled estimate $\varepsilon$ of the prediction error as a function of the training system size $n$. The $1nn$ Ising Hamiltonian with three disorder strengths is considered. The crosses represent the actual prediction errors $\overline{|\Delta_r u|}$ presented in Fig.~(\ref{scaling plots:1}). Notice that  $\varepsilon$ is multiplied by an appropriate scale factor to match the corresponding values of $\overline{|\Delta_r u|}$, allowing one to compare the scaling with $n$. 
 }
 \label{scaling plots:3}
\end{figure}

As mentioned above, the scaling of the prediction error $\overline{|\Delta_r u|}$ appears to relate to the decay of the normalized covariance $D(r)$. This suggests how to build an effective model that explains the observed scaling of $\overline{|\Delta_r u|}$ with $n$. 
We assume that the error originates from the discrepancy $\Delta D(r)=| D_l(r) -D_n(r)|$ between the training and the testing correlation functions, denoted here as $D_n(r)$ and $D_l(r)$.
We extrapolate $D_n(r)$ in the interval $[n/2,l/2]$ with $n$ and $l$ even numbers, by using a best fit with a stretched exponential $p(r)= a e^{-(r/b)^c}$ ($a,b,c$ are the fitting parameters), performed in the interval $[1,n/2]$ for $n \in \{12,18\}$ and $[2,n/2]$ for $n=24,64$.
The prediction error on a site $i$ is supposed to be due to contributions from all correlation discrepancies outside the training window $[i-n/2,i+n/2-1]$. The contribution for the distance $r$ is represented as a Gaussian random variable $\delta_r$ of zero mean and standard deviation $\Delta D(r)$,
    $\delta_r \sim \mathrm{N}(0,\Delta D(r) ).$
The total error $\chi_i$ results from the contributions outside the training window
\begin{equation}
    \chi_i =  \sum_{r=n/2}^{l/2-1} \delta_r + \sum_{r=n/2+1}^{l/2} \delta_r,
\end{equation}
where the two sums represent the contribution of the cutoff windows $[i+n/2,i+l/2-1]$ and $[i-l/2,i-n/2-1]$, respectively.
The final estimate of the prediction error is
\begin{equation}
    \varepsilon= \overline{\frac{1}{l}\sum_{i=1}^{l} \chi_i}.
\end{equation}
%
In Fig.~(\ref{scaling plots:3}), the scaling of the estimated error $\varepsilon$ is compared with the actual prediction errors $\overline{|\Delta_r u|}$, for the testing system size $l=64$, three  disorder strengths $h=1.8J,eJ,3.5J$, and training sizes $n \in \{12,24\}$. Notice that the goal is only to compare the scaling with $n$, and, in fact, $\varepsilon$ is scaled with an appropriate prefactor.
Interestingly, this  heuristic model is able to accurately capture the scaling of the prediction error with the training size $n$ for $h=1.8J$ and $h=eJ$. For $h=3.5J$ the estimation predicts a slightly faster decay in the error compared to the actual scaling. This discrepancy can be attributed to the increased difficulty in accurately reproducing a bijective map from $\mathbf{h}$ to $\mathbf{z}$  in the strong disorder regime, as discussed in Section~\ref{subsecvarying}. Indeed, this effect is not accounted for in this model.

\section{Conclusions}
\label{secconclusions}
We have shown that an accurate DFT approach for disordered quantum Ising chains can be implemented using scalable neural networks. These networks accurately map the transverse magnetization profile to the ground-state energy, and they allow finding the energy and magnetization of previously unseen realization of the random fields via a computationally efficient gradient-descent minimization of the predicted energy.
Notably, the scalable architecture allows predicting the properties of systems much larger than those used to train the network. This finding indicates a promising strategy to extend the reach of existing computational techniques which are accurate but limited to relatively small system sizes. Indeed, the prediction error in the thermodynamics limit  rapidly decreases as a function of the size $n$ used for training; specifically, the error approximately scales as $1/n^3$ in the worst case scenario, namely, at the quantum critical point, or even faster away from criticality.
Further investigations, which we leave for future studies, are due to establish the possible universality of the critical scaling.
Our analysis allowed us to shed light on the non-local character of the energy functional. Our DL-DFT approach systematically outperforms the purely local LDA and, thanks to increased kernel sizes and/or network depths, is able to capture long-range input-output functional dependencies.
It was also shown that, even when trained on relatively small system sizes, the DL functionals allow reconstructing the statistical correlations among input and output expectation values corresponding to very distant sites.
Finally, a heuristic model that describes the scaling of the prediction accuracy with the training size $n$ has been provided.
Future studies could also address higher dimensional systems or extend the DL approach to time-dependent DFT for quantum spin models~\cite{Tempel2012}. 
The resilience of DL functional to noisy training datasets should also be explored~\cite{saraceni2020scalable,ML_Q4}, as this could pave the way to learning DFTs from experimental measurements.

\section*{Acknowledgments} 
E.C. acknowledges hospitality in the QOQMS group led by A. Daley  at the University of Strathclyde. E.C. also acknowledges R. Connor from the QOQMS group and G. Scriva from the University of Camerino, for the interesting discussions and suggestions. 
S.P. and R.F. acknowledge support from  the PNRR MUR project PE0000023-NQSTI.
S.P also acknowledges support from the Italian Ministry of University and Research under the PRIN2017 project CEnTraL 20172H2SC4, and from the  CINECA award IscrC-NEMCAQS (2023), for the availability of high performance computing resources and support, as well as  from PRACE, for awarding access to the Fenix Infrastructure resources at Cineca, which are partially funded by the European Union’s Horizon 2020 research and innovation program through the ICEI project under the Grant Agreement No. 800858.

\appendix

\begin{figure}[h!]
\centering
\includegraphics[width=0.9\columnwidth]{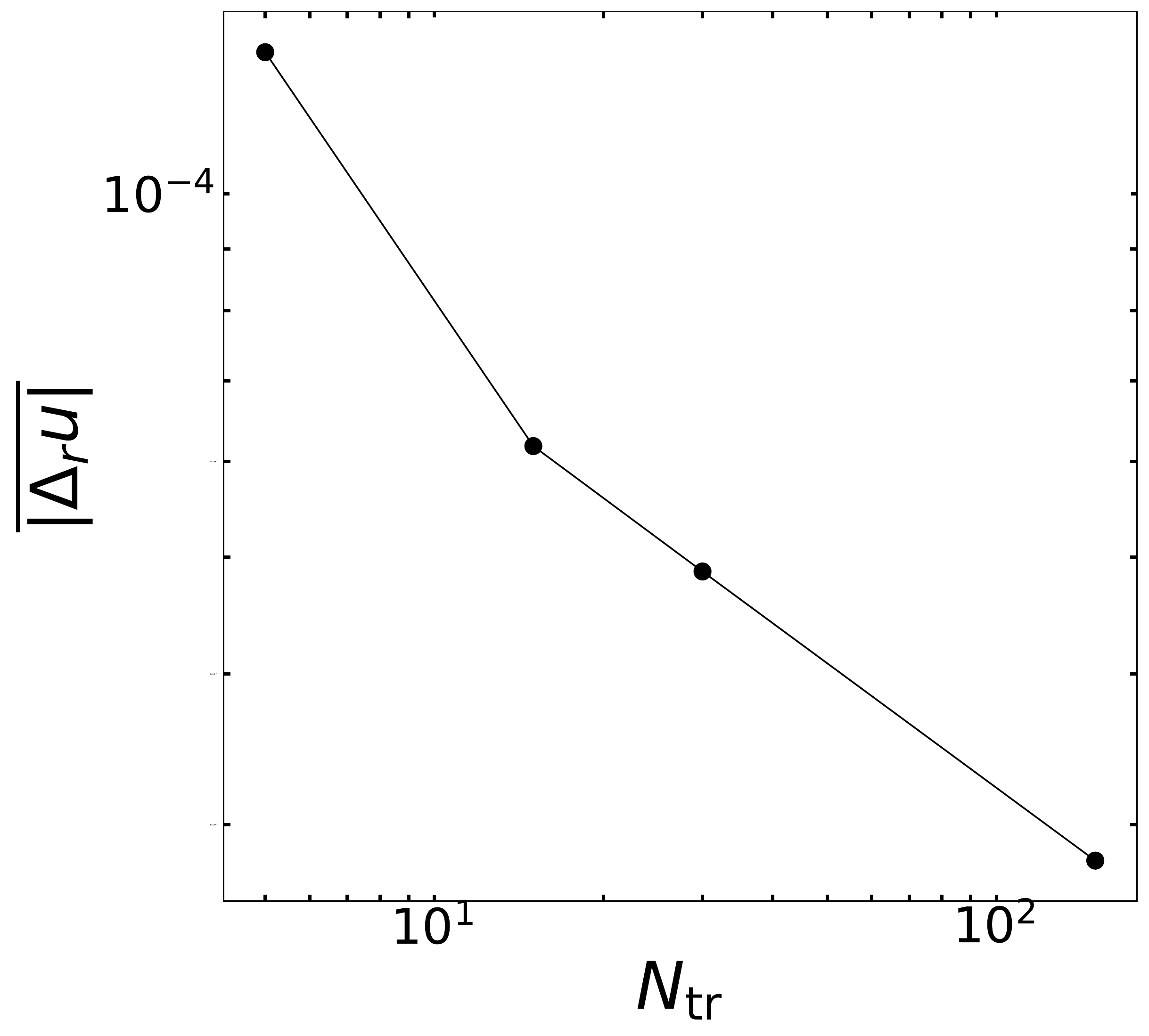}
\caption{Prediction error $\overline{|\Delta_r u|}$ as function of the number of instances in the training dataset $N_{\mathrm{tr}}$. 
The $1nn$ Hamiltonian at the critical point $h=eJ$ is considered, for a system size $l=16$.
The network features $N_{\mathrm{la}}=12$ layers.  }
\label{vs train size}
\end{figure}


\section{Systematic improvement of the prediction accuracy}
\label{section: systematic improvement}
\begin{figure}[h!]
\centering
\includegraphics[width=0.9\columnwidth]{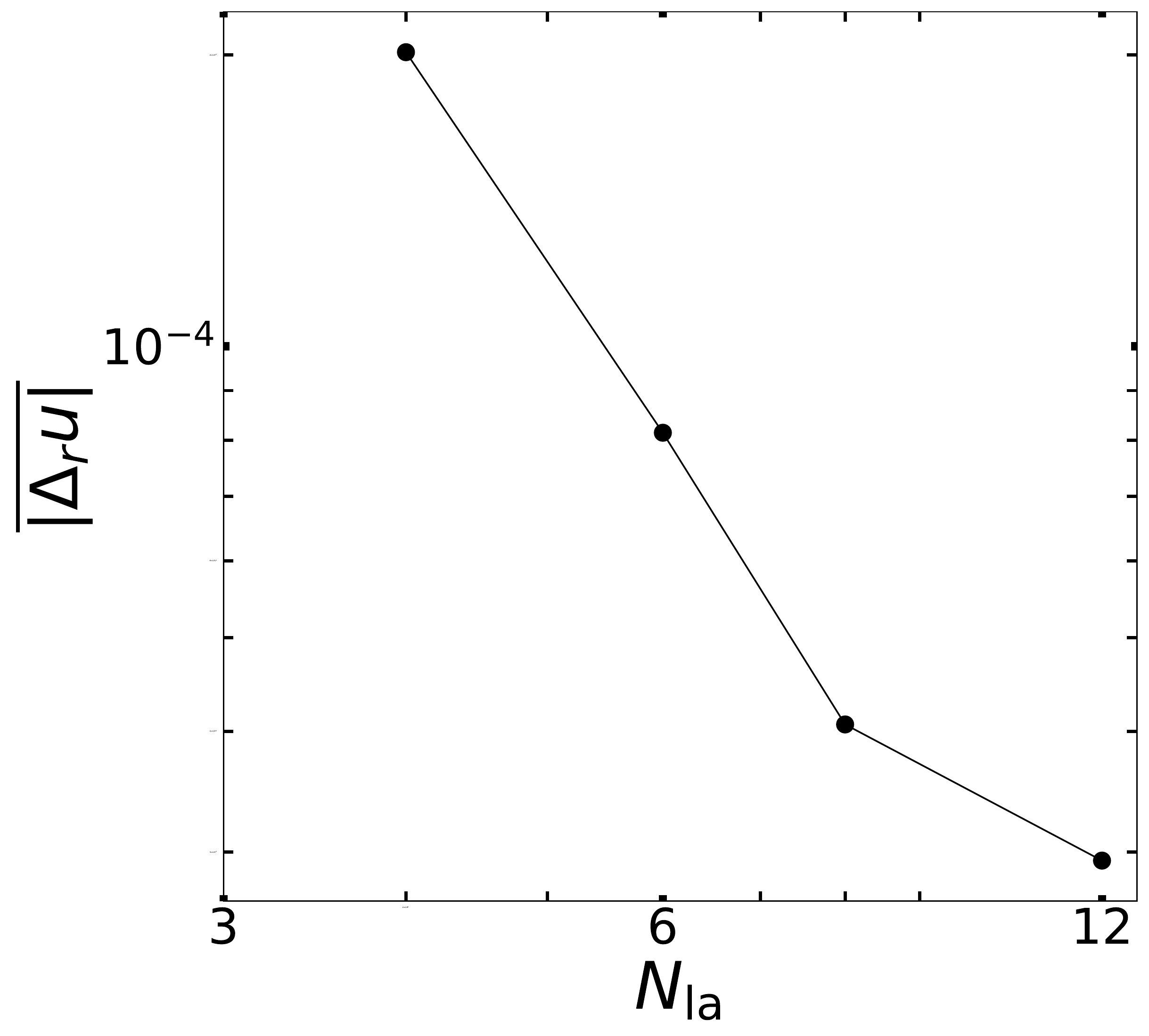}
\caption{Prediction error $\overline{|\Delta_r u|}$ as function of the number of layers $N_{l}$ in the neural network. The 1nn Hamiltonian at the critical point $h=eJ$ is considered, for a system size $l=16$. The size of the training dataset is $N_{\tr}=150000$.}
\label{vs layers}
\end{figure}

In Section~\ref{Results}, the accuracy of our DL functionals has been analyzed considering a fixed network structure and one training protocol. The error metrics were remarkably small. More important, perhaps, is that the error can be systematically controlled by increasing the network depth and/or the size of the training dataset. This is shown in Figs.~\ref{vs train size} and \ref{vs layers}. The 1nn Hamiltonian at the critical point $h=eJ$ is addressed. Only the prediction accuracy $\overline{|\Delta_r u|}$ is considered, since this metric also determines  the error $\overline{|\Delta_r e|}$ in the subsequent gradient-descent procedure. Specifically, we show that the prediction error rapidly decreases when the number of training instances $N_{\mathrm{tr}}$ increases, and also when the depth of the network $N_{\mathrm{la}}$ is augmented.

\section{Ferromagnetic phase transition in the next-nearest neighbor Ising chain}
\label{section: binder cumulant}
\begin{figure}[H]
\centering
\includegraphics[width=0.9\columnwidth]{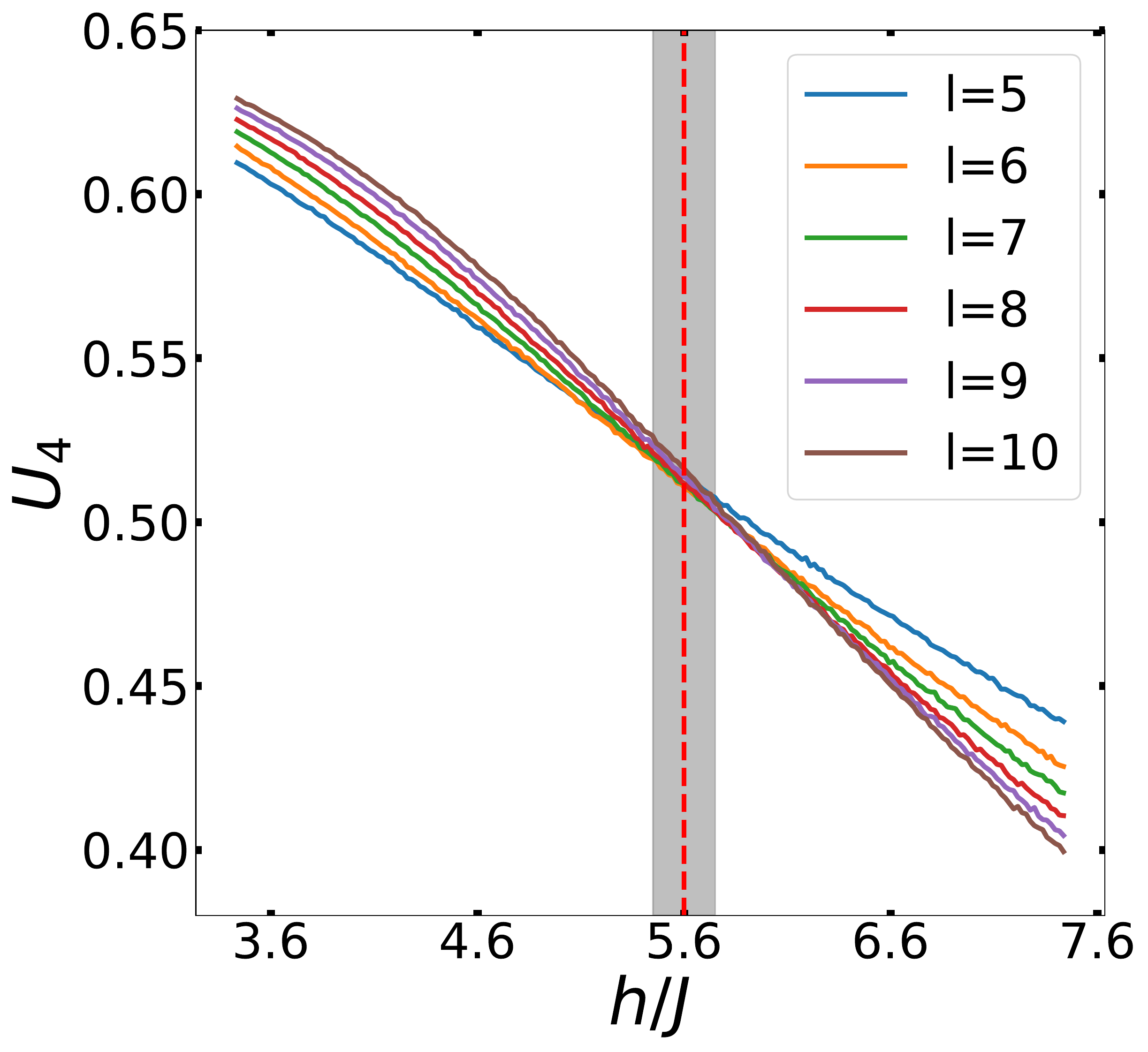}
\caption{Binder cumulant $U_4$ of the 2nn tranverse field Ising model with periodic boundary condition, as function of the disorder strength $h$, for different system sizes $l \in [5,9]$. The vertical gray bar indicates the approximate location of the ferromagnetic quantum phase transition.}
\label{binder cumulant}
\end{figure}

To determine the critical point of the ferromagnetic quantum phase transition in the $2nn$ Ising model, we perform a finite-size scaling analysis of the so-called Binder cumulant~\cite{Binder1981}, defined as:
\begin{equation}
     U_4= \overline{ \left(1 - \frac{\braket{X^4}}{3\braket{X^2}^2} \right)}.
\end{equation}
This is computed via exact diagonalization, as a function of the disorder strengths $h$ and for different chain lengths $l$. 
The disorder average is performed over $N_{\ts}=5\times10^4$ instances of the Hamiltonian. The results are shown in Fig.~\ref{binder cumulant}. The crossing of the datasets corresponding to different chain lengths denotes the critical point. This allows us to approximately estimate the critical disorder strength to be $h=5.60(15)J$.

\bibliography{main}{}

\end{document}